\documentclass[aps,prd,11pt,nofootinbib,tightenlines,twocolumn,superscriptaddress,notitlepage,preprintnumbers]{revtex4}
\usepackage{aas_macros}
\usepackage[a4paper,left=2cm,right=2cm, top=2.0cm,bottom=2.0cm]{geometry}
\usepackage[english]{babel}
\usepackage[applemac]{inputenc}
\usepackage[colorlinks,citecolor=blue,linktoc=all,linkcolor=black,urlcolor=black]{hyperref}
\usepackage{bm}
\usepackage{amsfonts,amsmath,amssymb}
\usepackage{graphicx}
\usepackage{booktabs, multirow}
\usepackage[detect-all]{siunitx}
\usepackage{hyperxmp}
\usepackage{color}

\newcommand{\Mh}{M_{\rm h}}
\newcommand{\Msun}{M_{\odot}}

\newcommand{\bh}{b_{\rm h}}
\newcommand{\bK}{b_{\rm K}}
\newcommand{\sigmav}{\sigma_{\rm v}}

\newcommand{\muk}{\mu_{\textit{\textbf{k}}}}

\makeatother

\begin{document}

\title{SHAPE: cosmology with cluster halo intrinsic alignments from subhalo distributions}

\author{Shogo Ishikawa}
\email{shogo.ishikawa.astro@gmail.com, shogo.ishikawa@yukawa.kyoto-u.ac.jp}
\affiliation{Department of Liberal Arts and Basic Sciences, College of Industrial Technology, Nihon University, Narashino, Chiba 275-8576, Japan}
\affiliation{Center for Gravitational Physics and Quantum Information, Yukawa Institute for Theoretical Physics, Kyoto University, Sakyo-ku, Kyoto 606-8502, Japan}
\author{Atsushi Taruya}
\affiliation{Center for Gravitational Physics and Quantum Information, Yukawa Institute for Theoretical Physics, Kyoto University, Sakyo-ku, Kyoto 606-8502, Japan}
\affiliation{Kavli Institute for the Physics and Mathematics of the Universe (WPI), University of Tokyo, Kashiwa, Chiba 277-8583, Japan}
\author{Takahiro Nishimichi}
\affiliation{Department of Astrophysics and Atmospheric Sciences, Faculty of Science, Kyoto Sangyo University, Motoyama, Kamigamo, Kita-ku, Kyoto 603-8555, Japan}
\affiliation{Center for Gravitational Physics and Quantum Information, Yukawa Institute for Theoretical Physics, Kyoto University, Sakyo-ku, Kyoto 606-8502, Japan}
\affiliation{Kavli Institute for the Physics and Mathematics of the Universe (WPI), University of Tokyo, Kashiwa, Chiba 277-8583, Japan}
\author{Teppei Okumura}
\affiliation{Institute of Astronomy and Astrophysics, Academia Sinica, No.~1, Section~4, Roosevelt Road, Taipei 10617, Taiwan}
\affiliation{Kavli Institute for the Physics and Mathematics of the Universe (WPI), University of Tokyo, Kashiwa, Chiba 277-8583, Japan}
\author{Satoshi Tanaka}
\affiliation{Center for Gravitational Physics and Quantum Information, Yukawa Institute for Theoretical Physics, Kyoto University, Sakyo-ku, Kyoto 606-8502, Japan}

\begin{abstract}
Galaxy clusters trace the most massive dark matter haloes, whose shapes and orientations reflect the imprint of the cosmic large-scale tidal field. 
This paper introduces the Subhalo-based Halo Alignment and Projected Ellipticity (SHAPE) technique, which reconstructs cluster halo shapes from the projected distribution of subhaloes, providing a novel approach to investigate intrinsic alignment (IA) correlations between cluster halo shapes and the surrounding density field. 
We measure halo shapes and orientations using different line-of-sight projection depths and find that, with modest projection depths, the shapes and orientations recovered by SHAPE show good agreement with those measured directly from the simulation particles. 
Using these SHAPE-derived shapes, we compute IA correlation functions from $N$-body simulations in both real and redshift space. 
The IA correlation multipoles exhibit features consistent with baryon acoustic oscillations around $100 \,h^{-1}{\rm Mpc}$ and show redshift-space distortion (RSD) effects that agree well with predictions from a non-linear alignment model incorporating RSD. 
We further demonstrate that the structure growth rate parameter can be robustly estimated without bias from these IA correlations, providing a new avenue for cosmological parameter estimation. 
Expanding the IA correlations in an associated Legendre basis yields results consistent with those from the standard Legendre expansion, but with improved statistical significance. 
These results suggest that SHAPE may enhance cosmological parameter constraints in future galaxy surveys. 
\end{abstract}

\preprint{YITP-25-59}
\maketitle

\section{Introduction} \label{sec:intro}
Galaxy clusters, the most massive gravitationally bound systems in the Universe, serve as natural laboratories for studying galaxy evolutions and star-formation activities \citep[e.g.,][]{gonzalez13,burke15,kravtsov18}. 
Moreover, they also play a crucial role in constraining cosmological parameters and understanding the structure formation of the Universe. 
This is because galaxy clusters form within highly biased density peaks of the background matter density field and possess deep gravitational potentials \citep{ps74,peebles80}, making them valuable probes for studying the underlying cosmology and large-scale structure \citep[e.g.,][]{bartelmann10,kravtsov12}. 
Recent extensive photometric galaxy surveys, i.e., the Hyper Suprime-Cam Subaru Strategic Program \citep[HSC SSP;][]{miyazaki18,aihara18}, the Kilo-Degree Survey \citep[KiDS;][]{dejong13}, and the Dark Energy Survey \citep[DES;][]{abbott16}, and have enabled the selection of a large number of galaxy clusters by covering huge survey volumes. 
As a result, galaxy clusters have rapidly gained attention as a powerful cosmological probe \citep[e.g.,][]{pacaud18,bocquet19,jaelani20,chiu20}. 

Shapes and orientations of galaxies and/or dark haloes exhibit correlation with their surrounding environments and such correlation is called intrinsic alignments \citep[IAs;][]{croft00,heavens00}. 
These alignments primarily arise from the linear tidal gravitational field induced by the large-scale structure, as described by the linear alignment (LA) model \citep{hirata04}. 
Initially, IAs were considered as unavoidable sources of systematic contamination in weak-lensing shear measurements \citep[e.g.,][]{joachimi15,mandelbaum18}. 
However, numerous theoretical studies have unveiled that IAs essentially contain valuable cosmological information. 
For instance, IAs can serve as tracers of the background tidal field, which in turn reflects the density field, revealing signatures of baryon acoustic oscillations \citep[BAOs;][]{chisari13,okumura19,vandompseler23} and the cosmic web \citep[][]{codis13,zhang22}, testing the gravitational theory \citep{chuang22,reischke22}, and constraining the geometry and dynamics of the Universe \citep{taruya20,okumura22}. 

Recent extensive galaxy redshift surveys have made significant progress in detecting the signals of IAs using observational data and extracting cosmological information from IA statistics, employing the aforementioned theoretical models \citep[e.g.,][]{singh15,okumura23,kurita23,vandompseler23,lamman24}. 
While most of the observed IA correlations are based upon galaxy shapes, it is widely recognized that IA signals from haloes are more prominent compared to galaxy IAs. 
This is because haloes are directly linked to the background density field and there exist various factors that influence galaxy shapes and their evolution through external physical mechanisms beyond gravitational tidal fields \citep[e.g.,][]{heavens00,lee11}. 
As a result, misalignments between galaxy shapes and halo shapes reduce the amplitudes of IA signals \citep{okumura09a}. 
Consequently, the IAs of haloes, rather than those of galaxies, directly provide notable information about the background gravitational fields and enable us to access cosmological information. 

Cluster haloes are highly suitable for measuring IA signals due to their strong correlation with the matter density field, resulting from their large gravitational potentials. 
While the small abundance of galaxy clusters in low-$z$  hinders apparent detections of IAs from cluster shapes, it has already been demonstrated that IA measurements of galaxy clusters in the intermediate-redshift Universe \citep[$z\lesssim0.6$;][]{vuitert17,vedder21} are feasible. 
Moreover, the IA analyses using mock cluster catalogues have served as powerful tools for extracting cosmological information \citep{shi23}.  

In this paper, we re-evaluate the detectability of IA signals in galaxy clusters by examining the subhalo distributions using a set of high-resolution $N$-body simulations. 
To accurately measure the IA signals of galaxy clusters using recent and upcoming extensive photometric surveys such as the HSC SSP and the Rubin Observatory's Legacy Survey of Space and Time \citep[LSST;][]{ivezic19}, it is crucial to qualitatively assess the impact of projection effects along the line-of-sight around the centers of cluster haloes, which can lead to dilution of the IA signals. 
By addressing the projection effects and demonstrating the effectiveness of this approach in detecting clear IA signals, including the BAO feature, we establish this method as the Subhalo-based Halo Alignment and Projected Ellipticity (SHAPE) technique. 

Unlike traditional IA analyses that rely on accurate measurements of individual galaxy shapes, the SHAPE method infers the overall cluster halo shape from the spatial distribution of subhaloes within galaxy clusters. 
This approach statistically mitigates the observational uncertainties and shape noize inherent to individual galaxy measurements, providing a stable and robust estimate of IA signals associated with the underlying tidal field. 
Moreover, because SHAPE directly uses the shapes of cluster-scale haloes, it probes the tidal field originating from large-scale structure more directly, rather than the local tidal interactions within individual haloes. 
Consequently, the SHAPE method offers a more physically direct measure of IAs induced by large-scale structure of the Universe. 

This paper is organized as follows. 
In Section~\ref{sec:nbody_simulation}, we describe the $N$-body simulations and subhalo catalogues used in this study. 
Section~\ref{sec:methodology} outlines the methods for measuring the shapes and orientations of cluster haloes using the SHAPE technique. 
The theoretical framework for IA statistics, including the LA and NLA models, is presented in Section~\ref{sec:ia_statistics}. 
Section~\ref{sec:validity_test} validates the SHAPE technique by comparing the IA signals measured in real space with those predicted from the fiducial cosmological parameters. 
In Section~\ref{sec:results}, we test the feasibility of constraining cosmological parameters by fitting the IA signals derived from the $N$-body simulations with the NLA model. 
Section~\ref{sec:discussion} discusses the implications of our results, and Section~\ref{sec:conclusions} provides the conclusions and summary of this study.

Throughout this paper, we adopt the cosmological parameters obtained from the analysis of the cosmic microwave background data by the Planck satellite assuming a flat-geometry $\Lambda$CDM model \citep{planck15}. 
Specifically, we use the following values: the cosmic density parameters $\Omega_{\rm m} = 0.3156$, $\Omega_{\rm \Lambda} = 0.6844$, and $\Omega_{\rm b} = 0.049$, the dimensionless Hubble parameter $h=0.6727$, the matter fluctuation averaged over a scale of $8$ $h^{-1}$Mpc $\sigma_{8} = 0.8322$, and the spectral index $n_{\rm s} = 0.9645$.
The mass of dark haloes is denoted as $\Mh$ and is expressed in units of $h^{-1}\Msun$. 

\section{Cosmological $N$-body simulations} \label{sec:nbody_simulation}
\subsection{Setup of the $N$-body simulations} \label{subsec:simulation_details}
We use datasets of $N$-body simulations that are carried out as a part of the Dark Quest II Project (Nishimichi et al. in prep.), which is a successor of the Dark Quest Project for developing a cosmological emulator \citep[DQ1;][]{nishimichi19,nishimichi21}. 
The $N$-body simulations are run using a cosmological TreePM code, Gravitational INtegrator for Kinematic Analysis of the darK Universe (\texttt{GINKAKU}; Nishimichi et al. in prep.), which is developed based on the FDPS framework\footnote{https://github.com/FDPS/FDPS} \citep{iwasawa16,Namekata_2018}, with the Partcle-Mesh extention from the \texttt{GreeM} code \citep{Yoshikawa_2005,Ishiyama_2009,Ishiyama_2012}. The calculation of tree forces is accelerated by the \texttt{Phantom-GRAPE} library~\citep{Tanikawa_2012,Tanikawa_2013} with the AVX-512 instructions~\citep{Yoshikawa_2018}.
The accuracy parameters are tuned such that the matter power spectrum is recovered within one percent compared to that from DQ1 based on \texttt{Gadget-2} \citep{Springel_2005}. 

In this work, we use the high mass-resolution dataset that employs $N_{\rm p} = 3,000^{3}$ particles with the comoving side length of the simulation box $L_{\rm box} = 1.0$ $h^{-1}$Gpc, corresponding to the mass resolution $3.23 \times 10^{9} h^{-1}\Msun$. 
Initial redshift is set to be $z_{\rm init} = 91.0$ and the linear power spectrum at the initial redshift is calculated using the cosmological Boltzmann code {\sc CLASS}\footnote{http://class-code.net/}  \citep{lesgourgues11}. 
We have three independent realizations with the same set up described above and make use of all of the realizations to evaluate the statistical errors of our IA measurements. 

\subsection{Subhalo catalogues} \label{subsec:subhalo_catalogue}
Subhalo catalogues from the simulation snapshots are generated using the publicly available halo finder {\sc ROCKSTAR}\footnote{https://bitbucket.org/gfcstanford/rockstar/src/main/} \citep{behroozi13}. 
Definition of halo masses used in this work is adopted the virial spherical overdensity \citep{bryan98} in the {\tt ROCKSTAR} halo catalogues. 
Haloes are identified from the snapshot at $z=0.319508$ by the friends-of-friends linking length $b=0.28$ and the force resolution is assumed $0.010$ $h^{-1}{\rm Mpc}$. 

To distinguish central haloes, which are gravitationally dominant and not located within the virial radius of a more massive halo, from subhaloes that reside within larger host haloes, we employ the {\tt FindParents} algorithm in the {\tt ROCKSTAR} code. 
In the subsequent section, we focus solely on measuring the shapes of central haloes. 

\section{Shape measurements of cluster haloes and catalogue constructions} \label{sec:methodology}
In this section, we present a methodology for measuring cluster halo IA signals. 
First, we introduce halo shape catalogues and density tracer catalogues derived from the subhalo catalogues generated in Section~\ref{subsec:subhalo_catalogue}. 
Subsequently, we outline the procedure for assessing halo shapes using the distributions of subhaloes within clusters. 
In this study, halo shapes are determined through projection onto the celestial sphere. 
As a result, we also explore the impact of projection on the shape measurements by varying projection depths. 

\subsection{Shape catalogue construction} \label{subsec:shape_catalogue}
\subsubsection{Shape and tracer catalogues} \label{subsubsec:shape_and_tracer}
As described in Section~\ref{subsec:subhalo_catalogue}, both the halo shape catalogues and density tracer catalogues are generated based upon the subhalo catalogues. 
The halo shape catalogues are constructed using the central haloes listed in the subhalo catalogues, with their shape information added after shape measurements using the density tracer catalogues described in Section~\ref{subsubsec:shape_measurements}. 
On the other hand, the density tracer catalogues consist of both central haloes and subhaloes, i.e., all haloes identified by the {\tt ROCKSTAR} halo finder, satisfying the following halo mass thresholds. 
Henceforth, for simplicity, we will refer to the halo shape catalogues and the density tracer catalogues as the shape catalogues and the tracer catalogues, respectively.  

We pose halo mass thresholds for both the shape and the tracer catalogues. 
Regarding the shape catalogues, the mass threshold is set to be $\Mh \geq 10^{13} h^{-1}\Msun$ to select galaxy groups and clusters with a sufficient number of member subhaloes for accurate shape measurements. 
On the other hand, the halo mass threshold for the tracer catalogues is set to be $\Mh \geq 10^{11} h^{-1}\Msun$, although our simulations can resolve much smaller haloes. 
This relatively conservative halo mass threshold for the tracer catalogues is intended to exclude haloes that lack galaxies, thus allowing us to measure the shapes of central haloes using observable galaxy distributions. 
Recent photometric observations have revealed that over half of haloes with masses of $\Mh \gtrsim 10^{11} h^{-1}\Msun$ possess galaxies satisfying the moderate magnitude thresholds at $z \sim 0.3$ \citep[e.g.,][]{zehavi05,zehavi11,coupon12,ishikawa20}. 
Consequently, our measurements of the shapes of cluster-scale haloes are comparable to those obtained by analysing observed member galaxies within galaxy clusters. 

\subsubsection{Cluster halo shape measurements} \label{subsubsec:shape_measurements}
\begin{figure}
\includegraphics[width=\columnwidth]{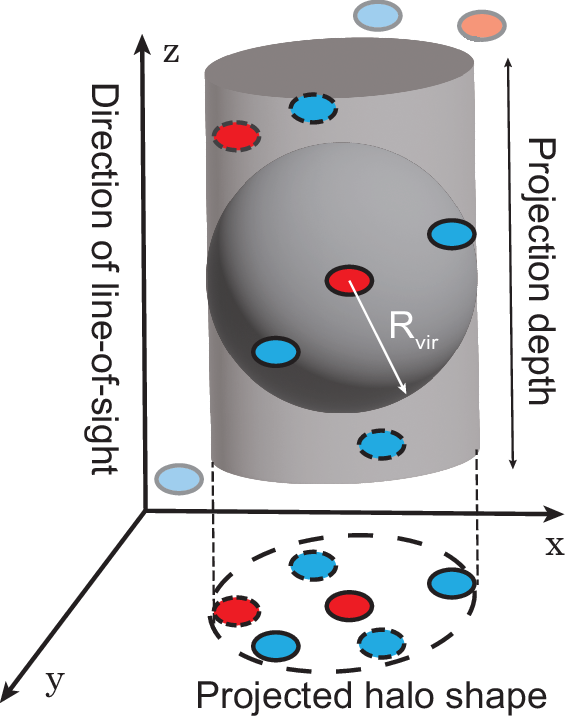}
\caption{A schematic illustration depicting the measurement of projected halo shapes using the SHAPE approach. In this diagram, the $x$-$y$ plane represents the celestial sphere, while the $z$-axis corresponds to the line-of-sight direction. The shape of each halo in our catalogue is determined using the spatial distribution of surrounding subhaloes and neighboring haloes within a cylindrical volume defined by a fixed projection depth along the line-of-sight. The virial radius, $R_{\rm{vir}}$, shown by the grey sphere, is used only as an initial guess for the halo boundary and does not restrict the subhalo selection in subsequent shape measurements. Thus, subhaloes and haloes located beyond $R_{\rm{vir}}$ but still within the cylindrical volume (blue and red ellipses enclosed by dashed lines) are also used in determining the halo shape. Specifically, member subhaloes gravitationally bound to the central halo (blue ellipses with solid black lines), other central haloes (red ellipses with dashed black lines), and non-member subhaloes (blue ellipses with dashed black lines) are all included, provided they lie within the cylinder. Haloes and subhaloes outside the cylindrical boundary (ellipses depicted in faint colors) are not used for shape determination. }
\label{fig:projected_shape}
\end{figure}

The shapes of cluster haloes measured in our analysis represent two-dimensional shapes projected onto the celestial sphere, although the three-dimensional positions of the subhaloes are available. 
This is because measuring halo shapes in three dimensions from observations is considerably challenging, even when spectroscopic data is available, due to the Fingers-of-God effect around cluster centers \citep{jackson72}. 
To estimate the shapes of galaxy clusters based on the spatial distributions of subhaloes, we consider a cylinder with a specific projection depth along the line-of-sight, centerd on a cluster halo listed in the shape catalogue \citep{reid09,okumura17}. 
Projected shapes of cluster haloes are then assessed using all the haloes enclosed within the cylinder. 
A schematic illustration of this process is presented in Figure~\ref{fig:projected_shape}. 

When evaluating the projected cluster shapes, we adopt the methods described in literature \citep[e.g.,][]{hoffmann14,knebe20}, which involve computing the second moment of inertia tensor $I_{ij}$. 
In computing the inertia tensor, we introduce a weight by inverse-squared distances from the centers of central haloes to the $n$-th subhalo $r_{n}$: 
\begin{equation}
I_{ij} = \sum_{n} \frac{x_{i, n} x_{j, n}}{r_{n}^{2}}, 
\label{eq:inertia}
\end{equation}
where $x_{i, n}$ and $x_{j, n}$ represent the $i$-th and $j$-th relative coordinate components of the $n$-th subhalo with respect to their central halo, corresponding to the $x$- and $y$-axes shown in Figure~\ref{fig:projected_shape}. 
By diagonalising the inertia tensor, we can derive the eigenvalues $\lambda_{1}$ and $\lambda_{2}$ ($\lambda_{1} > \lambda_{2}$), as well as the corresponding eigenvectors ${\bm v_{1}}$ and ${\bm v_{2}}$. 
The lengths of the major ($a$) and minor ($b$) axes can then be determined as $a = \sqrt{\lambda_{1}}$ and $b = \sqrt{\lambda_{2}}$, respectively. 
The orientations of the cluster haloes are defined by the angle of ${\bm v_{1}}$ with respect to one of the axes on the celestial sphere. 

We firstly specify the projection depth along the line-of-sight and then measure the shapes of cluster haloes listed in the shape catalogue. 
As with the case in literature \citep[e.g.,][]{lau12,suto16,suto17,harvey21}, we determine the shapes of the cluster haloes by iteratively computing the second moment of inertial tensor. 
The procedure of our cluster shape measurements is composed of three steps as follows. 
(i) First, we make an initial guess of the halo shape by assuming an elliptical shape. 
We select subhaloes listed in the tracer catalogue enclosed by a cylinder whose base is equal to the virial radius of the cluster halo, and which has a certain projection depth along the line-of-sight. 
For these selected subhaloes, we calculate the second moment of inertial tensor and measure the initial ellipsoidal shape of the cluster haloes. 
(ii) After determining the initial ellipsoidal halo shape, we then iterate by calculating the second moment of inertia tensor. 
For each iteration, we consider a new cylinder with a base defined by the projected halo shape obtained from the previous iteration and compare the axis ratio of the projected halo shapes to that derived in the previous iteration. 
(iii) If the change in the axis ratio from the previous iteration is less than $1\%$, we consider the halo shape to be converged. 
Subsequently, we add the final results of the halo shape and its orientation to the shape catalogue. 
If the iteration in step (ii) does not converge within $1,000$ times, we discard the central halo from the shape catalogue. 

We should keep in mind that our shape measurements are based upon the spatial distributions of subhaloes, not the distribution of dark matter particles. 
This can lead to an extremely flattened shape if the number of member subhaloes is quite small. 
To avoid encountering such cases, we refrain from measuring halo shapes if the number of member subhaloes drops below four in steps (i) and (ii). 
Furthermore, we maintain the initial areas of the projected halo shapes, as measured in step (i), when determining the halo shapes in each iteration of step (ii). 
This approach resolves the ambiguity in determining the absolute scale of the projected halo shapes, as the eigenvectors derived from the moment of inertia tensor only define directions and axis ratios of ellipses, not their absolute sizes. 
By fixing the ellipse area to that of the initial circular region defined by $R_{\rm vir}$, we avoid artificial expansion or contraction of halo shapes across iterations. 

In addition to generate shape catalogues in real space, we also construct the shape catalogues in redshift space. 
In redshift space, positions along the line-of-sight component $x_{k}^{\rm s}$ are shifted from those in real space $x_{k}^{\rm r}$ due to the redshift-space distortion as:
\begin{equation}
x_{k}^{\rm s} = x_{k}^{\rm r} + (1+z)\frac{{\bm v}_{k} \cdot \hat{x}_{k}}{H(z)}\hat{x}_{k}, 
\label{eq:rsd}
\end{equation}
where $H(z)$ is the Hubble parameter at redshift $z$, ${\bm v}_{k}$ represents the $k$-th component of the halo peculiar velocity in real space, and $\hat{x}_{k}$ is the unit vector along the $k$-th direction (corresponding to the line-of-sight direction, i.e., the $z$-axis in Figure~\ref{fig:projected_shape}). 
Note that we assume a plane-parallel approximation (it is so-called a distant observer approximation) for peculiar velocities of haloes. 

We have three realizations of the $N$-body simulations, all employing the identical setup described in Section~\ref{sec:nbody_simulation} (hereafter referred to as Realization~0, 1, and 2; see Table~\ref{tab:catalogue} for a summary of these realizations and their associated shape catalogues). 
Further details about the catalogues, including different projection depths, will be presented later in this section. 
These realizations yield three independent combinations of shape and tracer catalogues. 
Moreover, we possess particle snapshots of Realization~0 that enable us to measure the {\it true} halo shape as traced by its bound dark matter particles. 
In total, for each projection depth in both real space and redshift space, we have three sets of tracer and shape catalogues, with an additional {\it true} shape catalogue available for Realization~0. 

\subsection{Effects of the line-of-sight projection on shape measurements} \label{subsubsec:projection_depth}
Position measurements along the line-of-sight become uncertain in observations conducted in redshift space due to the influence of redshift-space distortions \citep{jackson72,kaiser87,hamilton98}. 
Consequently, it is essential to quantify the impact of this line-of-sight positional uncertainty on halo shape measurements by comparing the outcomes obtained in real space. 
Furthermore, the results produced by our SHAPE approach are affected by line-of-sight projections even in real space, as the number of contaminating haloes not bound by the central halo increases with larger projection depths. 
To comprehensively investigate the effects of line-of-sight uncertainties and projections within the SHAPE method, we generate multiple shape catalogues from a single realization by varying the projection depths along the line-of-sight. 

We generate halo shape catalogues by setting projection depths to $2.0$, $10.0$, $20.0$, $50.0$, $100.0$, $200.0$, and $500.0$ $h^{-1}{\rm Mpc}$  (corresponding to $\pm 1.0$, $\pm 5.0$, $\pm 10.0$, $\pm 25.0$, $\pm 50.0$, $\pm 100.0$, and $\pm 250.0$  $h^{-1} {\rm Mpc}$ from centers of central haloes), respectively, in both real and redshift space. 
The {\it true} halo shape catalogue, whose shapes are measured with the dark matter particle distribution, is also created only for Realization~0. 

\begin{figure}
\includegraphics[width=\columnwidth]{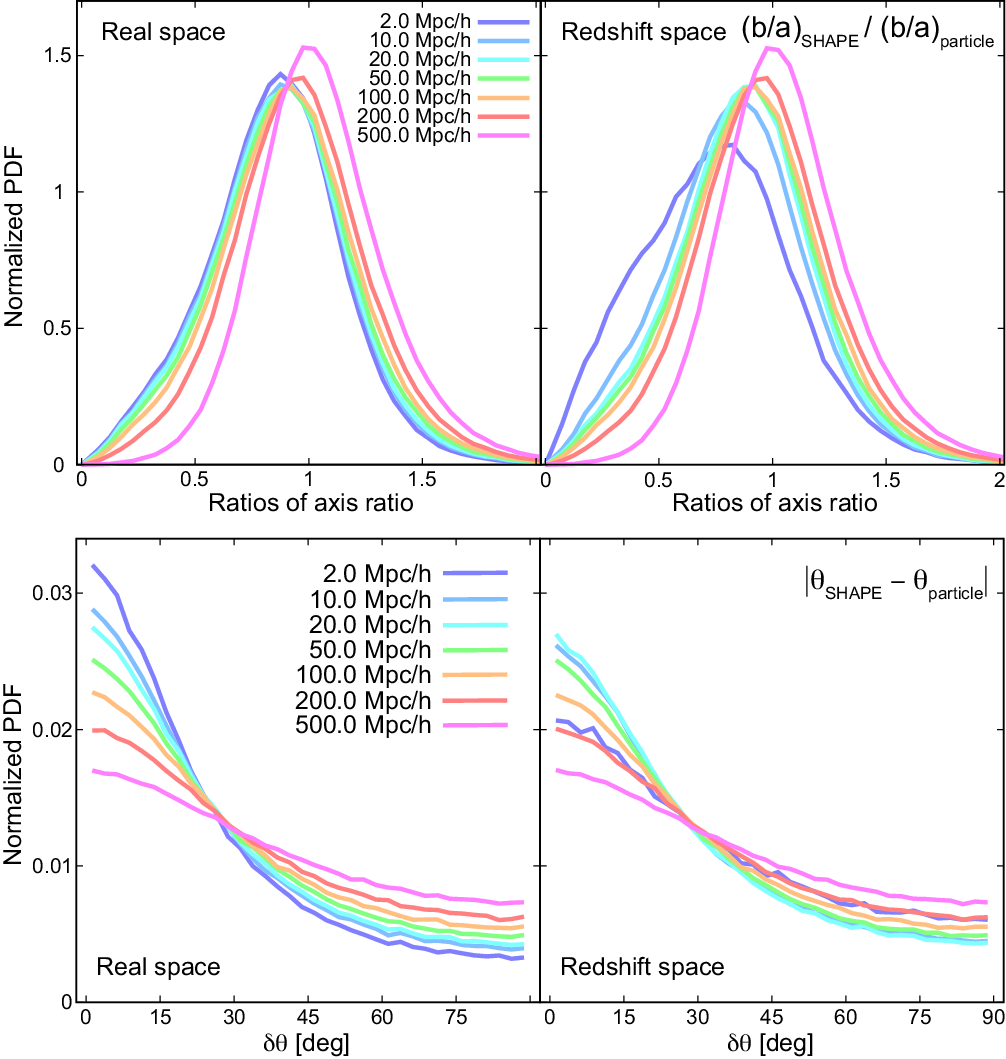}
\caption{The top panels of the plots show the axis ratio ratios of cluster halo shapes obtained using different projection depths, compared to the halo shapes measured by tracing dark matter particle distributions ({\it true} shape catalogues). 
The bottom panels display the major axis misalignments of the cluster halo shapes, again compared to the results of the {\it true} shape catalogues. 
The left panels depict the results measured in real space, whereas the right panels show the results in redshift space. 
The colors in the plots indicate the variation in the projection depth of the cylinders, which can be referred to in Figure~\ref{fig:projected_shape}. }
\label{fig:comp_shape}
\end{figure}

Figure~\ref{fig:comp_shape} illustrates the comparison of ratios of the axis ratio and misalignments of major axes between the haloes in each shape catalogue and those measured in the same haloes listed in the \textit{true} shape catalogues. 
As anticipated, the ratios of axis ratios are largely influenced by the projection depths in both real and redshift space. 
However, halo shapes tend to converge towards their original forms, i.e., the axis ratios measured using the SHAPE method approach those obtained from the {\it true} catalogue, as the projection depth increases, despite an increasing number of contaminating haloes included in shape measurements. 
This convergence occurs because cylinders with larger projection depths contain a significantly larger number of subhaloes gravitationally bound to the central haloes, while the contaminating subhaloes that are not gravitationally bound can be regarded as random sources of contamination that have minimal impact on the measured halo shapes. 
Notably, halo axis ratios in redshift space emphasize these trends: haloes with small projections are largely affected in their shapes due to the Fingers-of-God effect compared to those in real space, whereas axis ratios with large projection depths are identical to those in real space because they are almost free from redshift-space distortion effects. 
Comparing the ratios of axis ratios between real and redshift space, the results tend to be identical for the shape catalogues with relatively large projection depths ($\gtrsim 20.0$~$h^{-1}$Mpc). 

Furthermore, we investigate the effect of projection on the misalignments of major axes of the same haloes, with results shown in the lower panels of Figure~\ref{fig:comp_shape}. 
The misalignments in real space appear to yield intuitive results: haloes with small projection depths conserve the original orientation, whilst those with large projection depths have their angles randomly disturbed by contaminated haloes. 
The most striking result of this comparison is the misalignments in redshift space: haloes with small projection depths do not necessarily maintain their original orientations compared to those with moderate projection depths. 
This is because a large number of haloes with high peculiar velocities near the cluster centers escape from the cylinder due to the Fingers-of-God effect. 
As a result, the shapes of these cluster haloes with small projection depths become biased in a manner similar to haloes containing numerous randomly contaminating haloes at large projection depths. 

In summary, axis ratios of projected halo shapes are well reconstructed from their original shape by taking large projection depths in both real and redshift space, but orientations of major axes are largely affected by increasing projection depths. 
In redshift space, taking a small projection depth comparable to halo virial radii leads to significant misalignments from the original orientations due to the effects of contaminating subhaloes and the Fingers-of-God effect. 
To recover both the axis ratios and alignment angles of projected haloes in redshift space, it is necessary to choose moderate projection depths. 
These depths exceed the splashback radius, typically regarded as the physical boundary of haloes where material reaches its first apocenter after infall \citep[e.g.,][]{diemer14,more15}, but provide a practical observational scale. 
By minimising effects of the contamination and effectively recovering intrinsic halo properties, such projection depths strike a balance between observational constraints and theoretical consistency, making them well-suited for studies using the SHAPE technique. 

In the following analyses, we use shape catalogues with projection depths of $10.0$, $50.0$, and $100.0$~$h^{-1}$Mpc (corresponding to $\pm 5.0$, $\pm 25.0$, and $\pm 50.0$~$h^{-1}$Mpc from halo centers) to quantify the IA correlations in both real and redshift space. 
Smaller projection depths, such as $10.0$~$h^{-1}$Mpc, correspond roughly to spectroscopic observations with precise redshift measurements, whereas larger projection depths simulate observational scenarios using photometric redshifts that inherently involve significant uncertainties along the line-of-sight direction. 
We show the details of our shape and tracer catalogues in Table~\ref{tab:catalogue}. 

\begin{table*}
\caption{Details of the shape and tracer catalogues of each realization constructed in Section~\ref{subsubsec:projection_depth}. }
\label{tab:catalogue}
\begin{tabular}{lccccc}
\hline
Types of the catalogue & Realization & Projection depth & Threshold halo mass &\shortstack{The number of haloes\\in real space}  & $\bh^{a}$ \\
\hline
shape catalogue & $0$ &10.0~$h^{-1}$Mpc & $\Mh \geq 10^{13} h^{-1}\Msun$ & $341,634$ & -- \\
shape catalogue & $0$ &50.0~$h^{-1}$Mpc & $\Mh \geq 10^{13} h^{-1}\Msun$ & $374,380$ & -- \\
shape catalogue  &$0$ &100.0~$h^{-1}$Mpc & $\Mh \geq 10^{13} h^{-1}\Msun$ & $394,498$ & -- \\
{\it true} shape catalogue$^{b}$ & $0$ & -- & $\Mh \geq 10^{13} h^{-1}\Msun$ & $445,510$ & -- \\
tracer catalogue & $0$ & -- & $\Mh \geq 10^{11} h^{-1}\Msun$ & $37,542,634$ & $0.892^{+0.019}_{-0.018}$ \\
shape catalogue & $1$ &10.0~$h^{-1}$Mpc & $\Mh \geq 10^{13} h^{-1}\Msun$ & $342,728$ & -- \\
shape catalogue & $1$ &50.0~$h^{-1}$Mpc & $\Mh \geq 10^{13} h^{-1}\Msun$ & $375,633$ & -- \\
shape catalogue  &$1$ &100.0~$h^{-1}$Mpc & $\Mh \geq 10^{13} h^{-1}\Msun$ & $396,022$ & -- \\
tracer catalogue & $1$ & -- & $\Mh \geq 10^{11} h^{-1}\Msun$ & $37,547,182$ & $0.936^{+0.018}_{-0.018}$ \\
shape catalogue & $2$ &10.0~$h^{-1}$Mpc & $\Mh \geq 10^{13} h^{-1}\Msun$ & $342,419$ & -- \\
shape catalogue & $2$ &50.0~$h^{-1}$Mpc & $\Mh \geq 10^{13} h^{-1}\Msun$ & $375,144$ & -- \\
shape catalogue  &$2$ &100.0~$h^{-1}$Mpc & $\Mh \geq 10^{13} h^{-1}\Msun$ & $395,400$ & -- \\
tracer catalogue & $2$ & -- & $\Mh \geq 10^{11} h^{-1}\Msun$ & $37,536,067$ & $0.895^{+0.018}_{-0.018}$ \\
\hline
\multicolumn{6}{l}{\footnotesize$^a$ Halo biases of the tracer catalogues measured by the GG correlations. } \\
\multicolumn{6}{l}{\footnotesize$^b$ Shape catalogue measured using dark matter particle distribution. } \\
\end{tabular}
\end{table*}

\section{Intrinsic alignment statistics} \label{sec:ia_statistics}
\subsection{Correlation between halo shape and density field} \label{subsec:ia_statistics}
In this section, we introduce some quantities using our IA analysis. 
This study treats two types of anisotropic elliptical correlations in the IA analysis: the gravitational density--intrinsic ellipticity cross correlation (GI correlation) and the intrinsic ellipticity auto correlation (II correlation). 
See \citet{croft00}, \citet{heavens00}, \citet{hirata04}, and \citet{okumura20a} for more details. 

\begin{figure}
\includegraphics[width=\columnwidth]{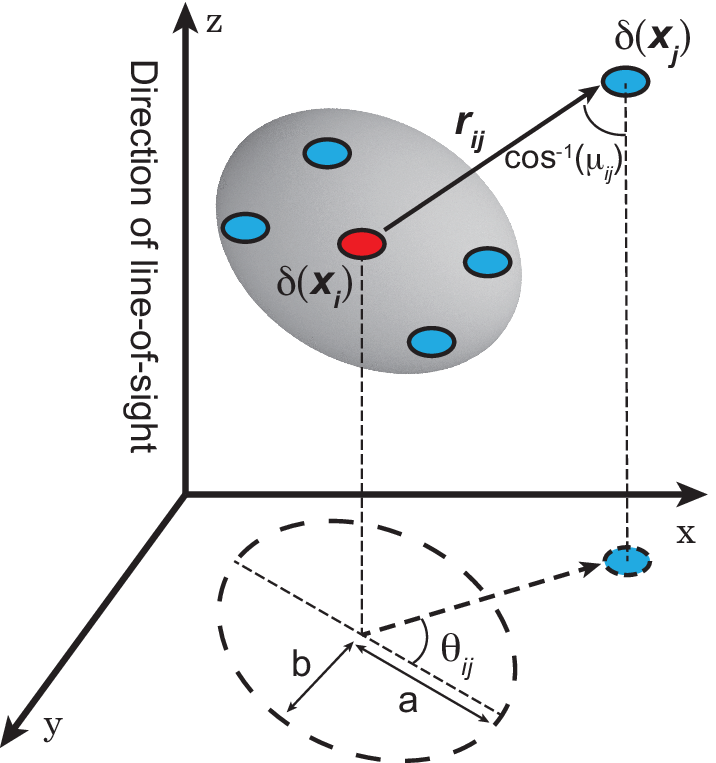}
\caption{A schematic illustration of the coordinate. 
In this illustration, the $z$-axis represents the direction of the observer's line-of-sight, and the $x$-$y$ plane corresponds to the celestial sphere. 
The density field of the $i$-th cluster listed in the shape catalogue is denoted as $\delta({\bm x}_{i})$, whereas $\delta({\bm x}_{j})$ represents the density field of the $j$-th subhalo (in the case of GI correlations) or cluster (II correlations). 
The three-dimensional distance between the $i$- and $j$-th objects is given by ${\bm r}_{ij}$, and the direction cosine between them is $\mu_{ij}$. 
The angle between the major axis of the projected shape of the $i$-th cluster and the direction vector between the $i$- and $j$-th objects is described as $\theta_{ij}$. 
The major and minor axes of the projected halo shape are represented by $a$ and $b$, respectively. }
\label{fig:coordinate}
\end{figure}

First, the ellipticity of the $i$-th galaxy cluster towards $j$-th object (density field or galaxy cluster) projected onto the celestial sphere can be expanded into two-components $\gamma_{(+, \times)}$:
\begin{equation}
\begin{pmatrix}
\gamma_{+} \\
\gamma_{\times} 
\end{pmatrix}
({\bm x}_i | {\bm x}_j) = \frac{1 - (b/a)_{i}}{1 + (b/a)_{i}}
\begin{pmatrix}
\cos(2\theta_{ij}) \\
\sin(2\theta_{ij})
\end{pmatrix}
,
\end{equation}
where the orientations and axis ratios of the cluster haloes are computed from the moment of inertia tensor defined by equation~(\ref{eq:inertia}). 
Here, $\theta_{ij}$ represents the angle between the direction vector from $i$-th cluster towards $j$-th object and the direction vector of the major axis of the $i$-th cluster shape, and $(b/a)_{i}$ is the axis ratio of the $i$-th cluster shape. 
We assume $(b/a)_{i} = 0$ throughout this paper \citep[see][for more details]{okumura09b,okumura23}. 
Figure~\ref{fig:coordinate} is a schematic diagram of this coordinate. 

Using the above ellipticity, the GI correlation $\xi_{\textrm{h}+}(\bm{r})$ can be express as:
\begin{equation}
1 + \xi_{\textrm{h}+}({\bm r}) = \langle  \gamma_{+}({\bm x}_{i}) (1 + \delta({\bm x}_{i})) (1 + \delta({\bm x}_{j}) ) \rangle,
\label{eq:GI}
\end{equation}
where ${\bm r}_{ij}$ represents the distance between the $i$-th cluster and $j$-th object, ${\bm r} = |{\bm x}_{j} - {\bm x}_{i}|$ and $\delta({\bm x}_{i})$ is the density fluctuation at a position ${\bm x}_{i}$. 
On the other hand, the II correlation can be written as:
\begin{equation}
1 + \xi_{++}({\bm r}) = \langle  \gamma_{+}({\bm x}_{i}) \gamma_{+}({\bm x}_{j}) (1 + \delta({\bm x}_{i})) (1 + \delta({\bm x}_{j}) ) \rangle.
\label{eq:II}
\end{equation}
In equation~\ref{eq:GI} and \ref{eq:II}, we can also evaluate correlations with the other combination, i.e., $\xi_{\times\times}$, by replacing $\gamma_{+}$ with $\gamma_{\times}$. 
Combining two components of the II correlation, $\xi_{++}$ and $\xi_{\times\times}$, we can additionally define two components in the II correlation as:
\begin{equation}
\xi_{\pm}({\bm r}) = \xi_{++}({\bm r}) \pm \xi_{\times\times}({\bm r}).
\label{eq:II_comb}
\end{equation}

In our definition, projected halo shapes are measured traced by subhalo distributions and it apparently induces anisotropies in the statistical quantities that include information of halo shapes both in real and redshift space. 
Furthermore, the elongated distributions around centers of massive clusters due to the Fingers-of-God effects in redshift space surely enhance the anisotropic distribution of subhaloes. 
Therefore, we can expand the multipole components of the GI and II correlations using the Legendre polynomials \citep{hamilton92} as follows:
\begin{equation}
\xi_{\ell}(r) = \frac{2 \ell + 1}{2} \int_{-1}^{1} d\mu \, \xi(r) \mathcal{P}_{\ell} (\mu_{ij}),
\label{eq:mulpole_integral}
\end{equation}
where $\ell$ is a multipole moment and $\mu_{ij}$ represents a direction cosine between $i$- and $j$-th objects. 
$\mathcal{P}_{\ell}$ represents the Legendre polynomial of order $\ell$.

\subsection{Estimators of GG and IA correlations} \label{subsubsec:estimators}
The halo auto correlations (gravitational density auto correlation; hereafter GG correlation) and IA correlations described in Section~\ref{subsec:ia_statistics} from the $N$-body simulations can be assessed using estimators of two-point statistics. 
The GG correlation function $\xi_{\rm hh}({\bm r})$ are measured using the Landy--Szalay estimator \citep{ls93} as follows:
\begin{equation}
\xi_{\rm hh} ({\bm r}) = \frac{{\rm DD} \left(\bm{r}\right) - 2{\rm DR}\left(\bm{r}\right) + {\rm RR}\left(\bm{r}\right)}{{\rm RR}\left(\bm{r}\right)}, 
\label{eq:ls}
\end{equation}
where DD$\left(\bm{r}\right)$, DR$\left(\bm{r}\right)$, and RR$\left(\bm{r}\right)$ represent the normalized data--data, data--random, and random--random pairwise counting with separation $\bm{r}$ within the periodic box, respectively. 

We use extended Landy--Szalay estimators that are weighted by halo shapes and orientations \citep{ls93,mandelbaum06,okumura19} as follows:
\begin{equation}
\xi_{\rm h+} (\bm{r}) = \frac{1}{\textrm{RR}(\bm{r})} \sum_{{\substack{i, j \\ \bm{r} = \bm{x}_{j} - \bm{x}_{i}}}} \gamma_{+}({\bm x}_i|{\bm x}_j), 
\label{eq:estimator_gi}
\end{equation}
and
\begin{equation}
\begin{split}
\xi_{\pm}(\bm{r})
&= \frac{1}{\mathrm{RR}(\bm{r})}
  \sum_{\substack{i,j\\\bm{r} =  \bm x_j - \bm x_i}}
  \Bigl[
    \gamma_{+}(\bm x_i \mid \bm x_j)\,\gamma_{+}(\bm x_j \mid \bm x_i) \\[-0.5ex]
&\quad\pm\;
    \gamma_{\times}(\bm x_i \mid \bm x_j)\,\gamma_{\times}(\bm x_j \mid \bm x_i)
  \Bigr].
\end{split}
\label{eq:estimator_ii}
\end{equation}
The number of shape--density, shape--shape, and random--random pairs used in the above estimators are normalized by their total number of pairs. 

The GG and IA correlation functions calculated from the $N$-body simulations can be expanded into their multipole components using the Legendre polynomials as follows:
\begin{equation}
\xi_{\rm X, \ell} (r) = \frac{2\ell + 1}{2} \sum_{{\substack{i, j \\ r = |x_{j} - x_{i}|}}} \xi_{\rm X}(r) \mathcal{P}_{\ell} (\mu_{ij}),
\label{eq:legendre_estimator}
\end{equation}
where ${\rm X}$ denotes the set of the correlations, i.e., ${\rm X} = \{\mathrm{hh, h+, +, -}\}$, specifically corresponding to the GG, GI, II~(+), and II~(-) correlations. 


\section{Testing the SHAPE approach} \label{sec:validity_test}
We have developed a methodology for assessing cluster halo shapes based on subhalo distributions, showcasing its potential for measuring IA correlations. 
In this section, we examine the validity of the SHAPE approach by (1) comparing IA correlations derived from $N$-body simulations with model predictions, and (2) quantifying the difference between the shape bias computed from the particle-based shape catalogue and the SHAPE-based shape catalogue to assess shape bias. 

\subsection{Comparison with model predictions} \label{subsec:comparison_with_model}
To begin, we calculate the GG and GI correlations in real space using the shape catalogue derived from our SHAPE technique and its corresponding tracer subhalo catalogue. 
For this validation, we employ a halo mass threshold of $\Mh = 10^{13} h^{-1}\Msun$ and a projection depth of $10.0$~$h^{-1}$Mpc when measuring the shapes of cluster haloes. 
It is noted that we performed similar tests for projection depths of $50\,h^{-1}{\rm Mpc}$ and $100\,h^{-1}{\rm Mpc}$, obtaining consistent results, further confirming the robustness of the SHAPE technique. 

The amplitudes of the GG and GI correlations in real space are governed by two free parameters: $\bh$ and $\bK$. 
We estimate these bias parameters to be approximately $\bh = 0.89$ and $\bK = -0.22$, values chosen to align with the amplitudes of the measured GG and GI correlation monopoles from our $N$-body simulation. 
Assuming the Planck~2015 cosmology \citep{planck15}, the linear growth rate at $z=0.319508$ is $f = 0.6936$. 


Following the NL RSD model, the GG and GI correlation functions in redshift space, $\xi_{X}^{s}(\bm{s})$, can be expressed as:  
\begin{equation}
\xi_{\rm X}^{s}(\bm{s}) = \int \frac{d^{3}\bm{k}}{(2\pi)^{3}} P_{\rm X}^{s}(\bm{k}) e^{i\bm{k}\cdot\bm{s}}. 
\label{eq:gg_gi_nrs}
\end{equation}
Here, $P_{\rm x}^{s}(\bm{k})$ denotes the GG \citep{scoccimarro04,taruya09,taruya10} and GI \citep{okumura24} power spectra in redshift space (see also \citep{taruya24} for an improved modeling based on perturbation theory calculations), defined as: 
\begin{equation}
\begin{split}
P_{\rm hh}^{s}(\bm{k})
&= \Bigl[
     \bh^{2}P_{\delta\delta}(k)
   + 2\bh\,f\,\mu_{\bm{k}}^{2}\,P_{\delta\Theta}(k) \\
&\quad
   + f^{2}\,\mu_{\bm{k}}^{4}\,P_{\Theta\Theta}(k)
  \Bigr]
  \;D_{\rm FoG}(\mu_{\bm{k}}\sigma_{v})\,,
\end{split}
\label{eq:pk_gg}
\end{equation}
and
\begin{equation}
\begin{split}
P_{\rm h+}^{s}(\bm{k})
&= \bK\,\frac{k_{x}^{2}-k_{y}^{2}}{k^{2}}
   \Bigl[
     \bh\,P_{\delta\delta}(k)
   + f\,\mu_{\bm{k}}^{2}\,P_{\delta\Theta}(k)
   \Bigr] \\
&\quad
   \; \times D_{\rm FoG}(\mu_{\bm{k}}\sigma_{v})\,.
\end{split}
\label{eq:pk_gi}
\end{equation}
where $k$ satisfies $k = |\bm{k}|$, and $\mu_{\bm{k}}$ is the direction cosine between the wavevector $|\bm{k}|$ and the observer's line-of-sight, defined as $\mu_{\bm{k}} = k_{z}/k$. 
$P_{\delta\delta}$ and $P_{\Theta\Theta}$ represent the non-linear auto power spectra of density field and velocity field, respectively, and $P_{\delta\Theta}$ describes their cross power spectrum \citep{hahn15}. 
In the linear theory limit, the auto and cross power spectra in density and velocity fields satisfy $P_{\delta\delta} = P_{\Theta\Theta} = P_{\delta\Theta}$. 
The function $D_{\textrm{FoG}}$ represents the Gaussian damping in power spectrum due to the Fingers-of-God effect \citep{peacock94,park94} described as:
\begin{equation}
D_{\rm FoG}(\mu_{\bm{k}}\sigma_{v}) = \exp{\left(-\frac{k^{2} \muk^{2} \sigma_{v}^{2}}{2} \right)},
\label{eq:fog}
\end{equation}
$\sigma_{v}$ is the non-linear velocity dispersion parameter in unit of $h^{-1}\textrm{Mpc}$. 
In the limit of $\sigma_{v} \rightarrow 0$, the Fingers-of-God damping becomes unity and the correlations in redshift space (equation~\ref{eq:gg_gi_nrs}) are close to those of in real space. 

\begin{figure}
\includegraphics[width=\columnwidth]{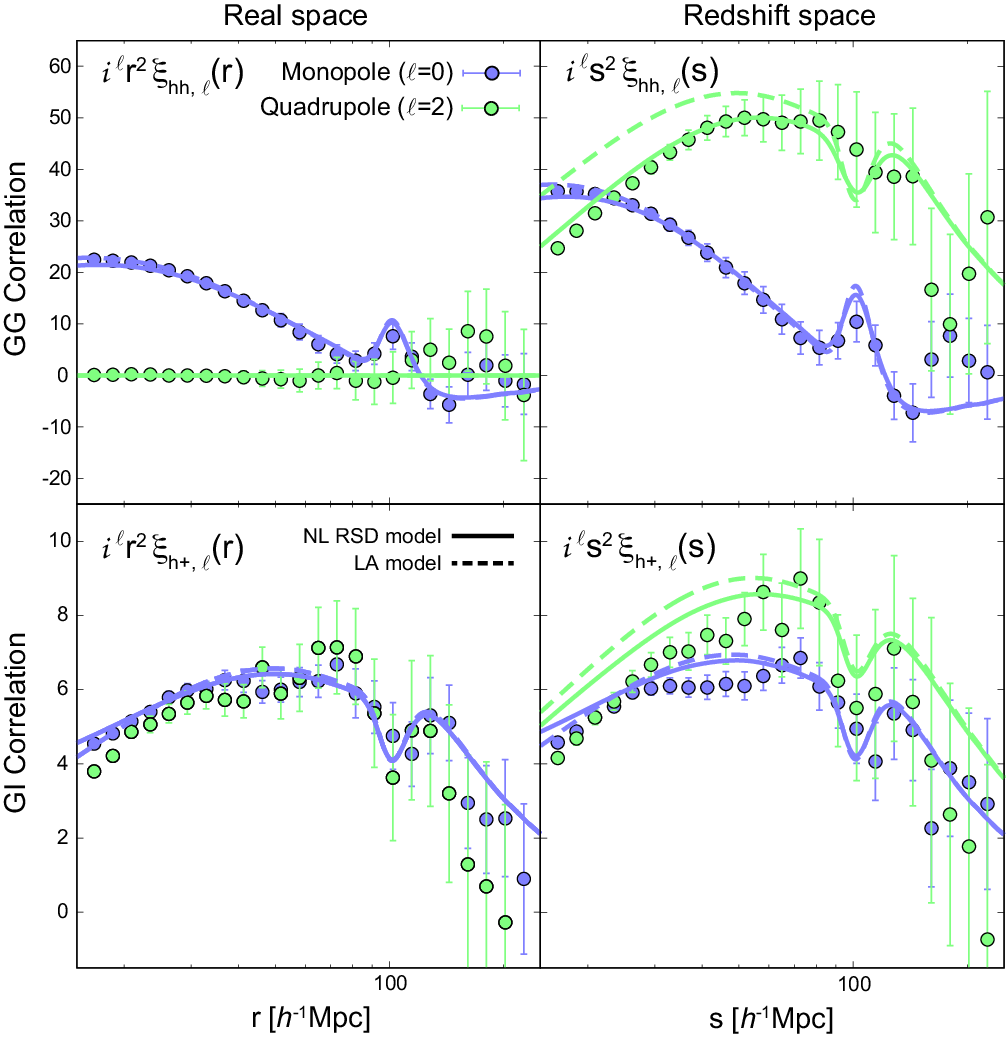}
\caption{GG and GI correlations of the monopole (blue) and the sign-inverted quadrupole (green) computed from the $N$-body simulations in Realization~2 (filled circles) and the NL RSD model \citep[solid lines;][]{taruya09,taruya10,okumura23}. The model-predicted correlations are calculated by assuming the parameters $\bh = 0.91$, $\bK = -0.22$, and $f = \sigmav = 0.0$ to represent the measured correlations from the $N$-body simulation at $z=0.319508$ in real space (left two panels). To check the validation of the SHAPE technique, we also derive the GG and GI correlations from the NL RSD model (solid lines) and linear-alignment (LA) model (dashed lines), assuming the same bias parameters but $\sigmav = 2.0$ and $f = 0.6936$, consistent with the prediction from the analytical model, and they are then compared with the correlations from the $N$-body simulations in redshift space (right two panels). It is noted that the model prediction of the GI correlations of the monopole and the quadrupole in real space entirely overlap. }
\label{fig:GG_and_GI}
\end{figure}
We present the GG and GI correlations measured from both the $N$-body simulations in Realization~2 and model predictions from the NL RSD model \citep{taruya09,taruya10,okumura24} and the LA model \citep[LA model,][]{catelan01,hirata04} in Figure~\ref{fig:GG_and_GI}. 
The GG and GI correlations in real space (left two panels in Figure~\ref{fig:GG_and_GI}) from the NL RSD model are calculated using the aforementioned halo and shape biases, with $f = \sigmav = 0.0$. 
In contrast, those in redshift space (right two panels in Figure~\ref{fig:GG_and_GI}) are evaluated assuming the same biases, but with $\sigmav = 2.0$ and $f = 0.6936$.

Notably, the model predictions computed to describe the GG and GI correlations in real space also successfully represent these correlations in redshift space from the $N$-body simulation, supporting the validity of the IA correlations derived from the SHAPE technique. 
Furthermore, the NL RSD model accurately predicts both the monopole and quadrupole correlations measured by the $N$-body simulations using the SHAPE technique. 
This implies that our SHAPE approach works well to characterize the IA correlation induced by the tidal field of large-scale structure. 

\subsection{Evaluation of shape bias} \label{subsec:shape_bias_ratio}
\begin{figure}  
\includegraphics[width=\columnwidth]{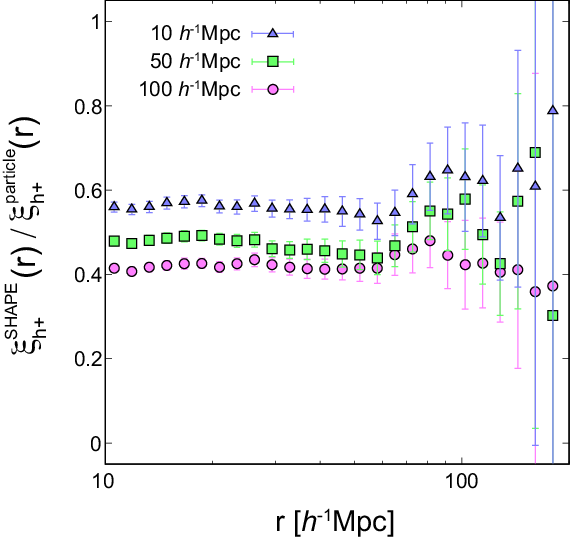}
\caption{Ratio of the GI correlation functions obtained from the SHAPE-based shape catalogue and the particle-based shape catalogue, $\xi^{\rm SHAPE}_{h+}(r) / \xi^{\rm particle}_{h+}(r)$, as a function of separation distance in real space. The different colors correspond to different projection depths used in the SHAPE approach: $10~h^{-1}$Mpc (upward triangles), $50~h^{-1}$Mpc (squares), and $100~h^{-1}$Mpc (circles). }
\label{fig:bk_ratio}  
\end{figure}

To assess the validity of the SHAPE approach, we compare the GI correlations in real space derived from the particle-based shape catalogue and the SHAPE-based shape catalogue. 
Figure~\ref{fig:bk_ratio} presents the ratio of the GI correlation functions, $\xi^{\rm SHAPE}_{\rm h+}(r) / \xi^{\rm particle}_{\rm h+}(r)$, measured in real space. 
The GI correlations from the SHAPE approach are derived using three shape catalogues with the projection depths of $10$, $50$, and $100$ $h^{-1}$Mpc. 

The results indicate that, within the error bars, the shape bias ratio remains nearly constant across all scales. 
This suggests that the GI correlation obtained via the SHAPE approach is broadly consistent with that from the conventional particle-based IA method, aside from a reduction in amplitude likely caused by shape noise. 
Furthermore, we find that the shape bias ratio systematically decreases as the projection depth increases, implying that deeper projections introduce additional noise. 
One plausible explanation is that including a larger line-of-sight volume draws in subhaloes that are not genuinely bound to the target cluster, thus diluting the intrinsic alignment signal and lowering the overall amplitude of the measured GI correlation. 

Nevertheless, we identify a small bump around \(\sim 100\,h^{-1}\mathrm{Mpc}\), which appears most prominently for shape catalogues with shallow projection depths (e.g.,\ \(10\,h^{-1}\mathrm{Mpc}\)). 
By contrast, in the \textit{true} catalogue the BAO signature manifests as a pronounced trough at \(\sim 100\,h^{-1}\mathrm{Mpc}\). 
Hence, the GI correlations from catalogues with small projection depths exhibit a shallower BAO trough than in the \textit{true} case. 
This discrepancy can be attributed to the fact that shallow projection depths primarily capture subhaloes near the cluster center, accurately reproducing the central alignment but underestimating contributions from the cluster outskirts. 
The missing outer regions play a key role in reinforcing the large-scale alignment, particularly on BAO scales. 
Conversely, as the projection depth increases, more of the physically associated subhaloes in the outskirts are recovered, enabling a deeper BAO trough that better matches the \textit{true} catalogue. 
At the same time, noise from foreground/background contamination also increases, which reduces the overall amplitude of the GI correlation but leaves the BAO feature more faithfully reproduced. 

These findings confirm that the SHAPE approach effectively captures IA correlations while accounting for shape noise. 
Moreover, our results highlight that a sufficiently large projection depth better recovers the BAO trough by incorporating the full halo outskirts, even though this can introduce additional noise and lower the overall GI amplitude. 
Thus, when probing BAO-scale features, choosing a deeper projection depth may offer a more faithful representation of the underlying halo-large-scale structure alignment. 

\section{Cosmological inference from the SHAPE approach} \label{sec:results}
In this section, we perform the parameter inference to further test our SHAPE approach. 
To be precise, we compare the GG and SHAPE statistics measured from $N$-body simulations with the analytical models, allowing the linear growth rate $f$ to be free. 
We use shape catalogues with projection depths of $10.0$, $50.0$, and $100.0$~$h^{-1}$Mpc to quantitatively investigate the effect of line-of-sight projection depth on IA statistics. 

\subsection{MCMC analysis with theoretical model} \label{subsec:model_fitting}
We present the fitting procedure of the GG and IA correlations obtained from the $N$-body simulations using the analytical model of the non-linear correlations with the redshift-space distortion effect \citep[NL RSD model;][]{taruya09,taruya10,okumura24}. 
The NL RSD model was developed based upon the LA model that connects the observed shape field to the background gravitational tidal field {catelan01,hirata04} and its expansion to the non-linear regime known as the non-linear alignment model \citep[NLA model,][]{bridle07}. 
In the NL RSD model, the II correlations in redshift space $\xi_{\pm, \mathrm{model}}^{s}(s)$ can be calculated using the following power spectrum as:
\begin{equation}
\begin{split}
P_{\pm}^{s}(\bm{k})
&= \bK^{2}\,
   \frac{(k_{x}^{2} - k_{y}^{2})^{2} \pm (2k_{x}k_{y})^{2}}{k^{4}}
   \;P_{\delta\delta}(k) \\[-0.5ex]
&\quad\times
   D_{\mathrm{FoG}}(\mu_{\bm{k}}\sigma_{v})\,. 
\end{split}
\label{eq:pk_ii}
\end{equation}
and transform it into configuration space using equation~\ref{eq:gg_gi_nrs}. 

In the NL RSD model, as mentioned in Section~\ref{sec:validity_test}, we have four free parameters: the halo bias $\bh$, the shape bias $\bK$, the linear growth rate $f$, and the velocity dispersion of haloes $\sigmav$ (equation~\ref{eq:pk_gg}, \ref{eq:pk_gi}, and \ref{eq:pk_ii}. 
Theoretical templates from the NL RSD model in redshift space are computed by varying the four free parameters, sampled using the MCMC technique with the \texttt{emcee} package\footnote{https://emcee.readthedocs.io/en/stable/} \citep{foreman13}. 
The parameters are constrained through maximum likelihood estimation with flat priors as follows:
\begin{equation}
\begin{split}
\ln{\mathcal{L}} \;=\;& -\frac{1}{2} \sum_{\ell, \ell'} \sum_{\mathrm{X}, \mathrm{X'}} \sum_{i, j} 
\Bigl[\xi_{\mathrm{X}, \ell}^{S}(r_i)\;-\;\xi_{\mathrm{X}, \ell, \mathrm{model}}^{S}(r_i)\Bigr] \\
&\times \,C^{-1}_{ij ,\mathrm{XX'}, \ell\ell'}\,
\Bigl[\xi_{\mathrm{X'}, \ell'}^{S}(r_j)\;-\;\xi_{\mathrm{X'}, \ell', \mathrm{model}}^{S}(r_j)\Bigr],
\end{split}
\label{eq:fitting_model}
\end{equation}
where $\xi_{\mathrm{X},\ell}^{S}(r_i)$ and $\xi_{\mathrm{X}, \ell, \mathrm{model}}^{S}(r_i)$ are respectively the measured and model-predicted correlation functions in redshift space for the $i$-th radial bin. 
The indices $\mathrm{X}$ and $\mathrm{X'}$ run over $\{\mathrm{hh,\,h+,\,+,\,-}\}$, while $\ell$ and $\ell'$ take values in $\{0,\,2,\,4\}$, corresponding to the monopole, quadrupole, and hexadecapole components expanded by Legendre polynomials. 
Here, $C^{-1}_{ij,\mathrm{XX'},\ell\ell'}$ denotes the inverse of the full covariance matrix for all correlation functions, with $i$ and $j$ indexing the radial bins. 
In calculating the correlation functions, the non-linear matter power spectra are generated using the publicly available cosmological Boltzmann code {\sc CLASS} \citep{lesgourgues11}, which incorporates the revised {\sc halofit} model from \citet{takahashi12}. 
The resulting power spectra for the GG and IA correlations are then transformed into configuration space via the FFTLog algorithm \citep{talman78,hamilton00}, implemented through the publicly available {\sc HANKL} code. package\footnote{https://hankl.readthedocs.io/en/latest/} \citep{karamanis21}. 
The covariance matrices of each correlation function are evaluated by the jackknife resampling method with dividing the periodic box into $512$ sub-box. 
The Hartlap factor \citep{hartlap07} is taken into account for evaluating the inverse matrices from the covariance matrices by the jackknife resampling. 
We use the GG and IA correlations at scales ranging from $20.0$~$h^{-1}$Mpc to $200.0$~$h^{-1}$Mpc in the fitting procedures to quantify the applicable scale  of the SHAPE technique. 
The mean posteriors are calculated after the MCMC fittings of $30,000$ steps and $15$ walkers, including $30\%$ burn-in phases for each projection depth. 

We simultaneously fit the GG, GI, II~(+), and II~(-) correlations expanded using the Legendre polynomials ($12$ multipole components in total for each shape catalogue) measured from shape catalogues with the same projection depth in redshift space by assuming the NL RSD model. 
Correlation functions in real space are also calculated using the same $\bh$ and $\bK$ parameters that are determined from the MCMC fitting in redshift space with setting $f = \sigmav = 0.0$. 
This corresponds to neglecting the redshift-space distortions, i.e., the Finger-of-Gods effect \citep{jackson72} and the Kaiser effect \citep{kaiser87}, in the real-space model. 
Besides, the correlation functions derived from the LA model both in real and redshift space are computed using the same parameter sets constrained by the NL RSD model in redshift space for comparison. 

\subsection{Results} \label{subsec:results}
\subsubsection{GG-only fitting} \label{subsubsec:result_gg_only}
Before doing a joint fitting analysis with SHAPE-derived  IAs, let us first perform the analysis using the GG correlation alone (see Section~\ref{subsubsec:result_gg+ia} for the GG+IA joint fitting analysis). 
In the fitting procedure, we set $\mathrm{X} = \mathrm{X'} = \{\mathrm{hh}\}$ in equation~\ref{eq:fitting_model}. 

The results of the GG-only fitting are presented in Table~\ref{tab:ia_fitting}. 
Focusing on the constraints on the cosmic linear growth rate $f$, only Realization~2 successfully recovers the Planck~2015 prediction of $f=0.6936$ within the $1\sigma$ confidence level among the three realizations. 
The other two realizations exhibit deviations ranging from $1.1\sigma$ to $2.6\sigma$ from the fiducial value. 

\subsubsection{GG + SHAPE-derived IA fitting} \label{subsubsec:result_gg+ia}
\begin{figure*}
\includegraphics[width=18cm]{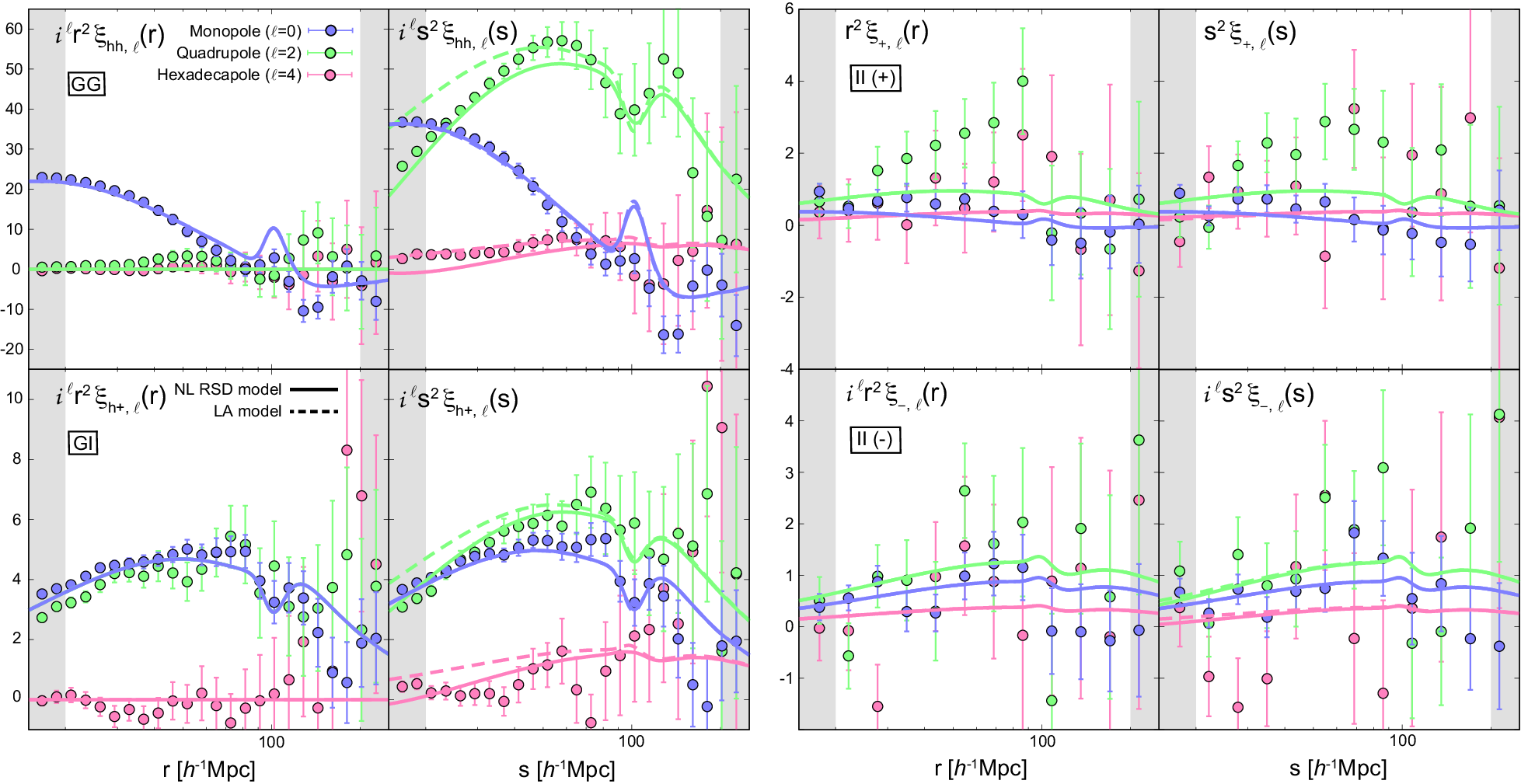}
\caption{Multipole components of the GG and SHAPE-derived IA correlation functions expanded using Legendre polynomials. Filled circles represent measurements from the $N$-body simulations obtained via the SHAPE method, where different colors denote different multipole orders: monopole ($\ell=0$; blue), quadrupole ($\ell=2$; green), and hexadecapole ($\ell=4$; red). Solid curves indicate the best-fitting NL RSD model predictions, whereas dashed lines correspond to the LA model predictions calculated using parameters derived from the NL RSD model fits. Results in real and redshift spaces are shown side-by-side for clarity. Grey-shaded regions represent scales excluded from the fitting analysis (${\rm s}<20\,h^{-1}{\rm Mpc}$ and ${\rm s}>200\,h^{-1}{\rm Mpc}$). 
Note that the quadrupole ($\ell=2$) components for GG, GI, and II~(-) correlations have inverted signs to enhance visual clarity. Additionally, the II~(+) and II~(-) correlations shown here have bin widths doubled relative to the GG and GI correlations for improved visibility; however, the actual fitting procedures employ identical bin widths for all correlation functions. }
\label{fig:gi-ii}
\end{figure*}

We then perform a joint fitting of the GG and SHAPE-derived IA correlations using the NL RSD model. 
The fitting results for all shape catalogues (projection depths of $10$, $50$, and $100\,h^{-1}{\rm Mpc}$ across the three realizations) are also summarized in Table~\ref{tab:ia_fitting}. 
Figure~\ref{fig:gi-ii} displays the GG and SHAPE-derived IA correlations measured from the shape catalogue with a projection depth of $100\,h^{-1}{\rm Mpc}$ in Realization~0, along with their best-fitting predictions obtained from the NL RSD model. 
It is noted that the II~(+) and II~(-) correlations shown in Figure~\ref{fig:gi-ii} have bin widths doubled relative to the GG and GI correlations for improved visibility; however, the actual fitting procedures employ identical bin widths for all correlation functions. 

What is most striking about Figure~\ref{fig:gi-ii} is that the monopole of the GI correlation measured using the SHAPE technique clearly exhibits the BAO trough at $\sim 100\,h^{-1}\mathrm{Mpc}$. 
Notably, the BAO feature becomes more diluted for projection depths of $10$ and $50\,h^{-1}\mathrm{Mpc}$. 
This suggests that larger projection depths may help enhance the correlation between cluster halo shapes and the background tidal field, as discussed in Section~\ref{subsec:shape_bias_ratio}. 
Furthermore, the monopole and quadrupole of the GI correlation in redshift space show clearly distinct amplitudes and shapes, explicitly highlighting the impact of RSD effects on the GI correlation. 

The II~(+) and II~(-) correlations, although exhibiting relatively lower signal-to-noise ratios compared to the GI correlation, still provide valuable complementary information, improving the constraints on cosmological parameters. 
To explicitly evaluate their impact, we conducted MCMC fits excluding the II correlations, i.e., using $\mathrm{X} = \mathrm{X'} = \{{\mathrm{hh}, \mathrm{h+}}\}$ in equation~\ref{eq:fitting_model}. 
When excluding the II correlations with a $100\,h^{-1}\mathrm{Mpc}$ projection depth, the inferred linear growth rate parameters are $f = 0.725$, $0.646$, and $0.680$ for Realizations~0, 1, and 2, respectively, compared to $f = 0.697$, $0.659$, and $0.725$ with the II correlations. 
Given the fiducial value of $f=0.6936$, the inclusion of the II correlations improves the overall accuracy and robustness of the inferred growth rate parameter, highlighting the significance and utility of incorporating the II correlations in cosmological analyses. 

We now turn our attention to the differences arising from varying projection depths of $10.0$, $50.0$, and $100.0\,h^{-1}\mathrm{Mpc}$, and assess their impact on cosmological parameter estimation and the goodness-of-fit. 
Overall, the inferred cosmic linear growth rate parameter, $f$, exhibits a consistent trend towards improved agreement with the fiducial value of $f=0.6936$ by the Planck~2015 results as the projection depth increases. 
For instance, in Realization~0, the measured values of $f$ progressively improve from $0.702^{+0.027}_{-0.026}$ at $10.0\,h^{-1}\mathrm{Mpc}$ to $0.697^{+0.026}_{-0.026}$ at $100.0\,h^{-1}\mathrm{Mpc}$. 
Similar improvements are also observed in the other realizations, highlighting that larger projection depths contribute positively to the robustness and accuracy of cosmological constraints derived from the SHAPE technique. 

Furthermore, we assess the goodness-of-fit by examining the reduced $\chi^{2}$ values across different projection depths. 
For Realization~0, the reduced $\chi^{2}$ value decreases notably from $0.832$ at a projection depth of $100.0\,h^{-1}\mathrm{Mpc}$ to $0.676$ at $10.0\,h^{-1}\mathrm{Mpc}$. 
A reduced $\chi^{2}$ substantially below unity, as seen at smaller projection depths, may suggest potential over-fitting, indicating that statistical fluctuations in the data might be excessively captured by the model. 
Conversely, larger projection depths yield reduced $\chi^{2}$ values closer to unity, indicating that the NL RSD model provides a more statistically balanced and realistic representation of the measured correlations at greater projection depths. 

Taken together, these results suggest that employing deeper projection depths enhances both the precision of cosmological parameter estimation and the statistical consistency of model fits. 
Thus, the SHAPE technique, when applied with sufficiently large projection depths, can reliably recover cosmological parameters with improved accuracy and robustness. 

\begin{table*}
\caption{Posterior mean and best-fitting values of free parameters of the NLA model fitted by the GG-only and GG + SHAPE-derived IA correlations expanded using standard Legendre polynomials for each realization in redshift space. Values in parentheses indicate the best-fitting values during the MCMC fitting. Assuming the cosmological parameters of Planck~2015 observation, the fiducial value of $f$ at $z=0.319508$ is $f=0.6936$. Note that, for all cases, the values of $\chi^{2}_{\rm total}/{\rm dof}$ for the posterior mean and best-fitting results are identical at the given precision (three decimal places). We therefore report only a single value. }
\label{tab:ia_fitting}
\resizebox{\textwidth}{!}{%
\begin{tabular}{clccccc}
\hline
Realization & Projection$^a$ & $\bh$ & $\bK$ & $f$ & $\sigma_{v}$$^b$ & $\chi^{2}_{\rm total}$ / dof$^{c}$ \\
\hline
\multicolumn{7}{c}{GG-only fitting} \\
\hline
$0$ & -- & $0.870^{+0.018}_{-0.018}\,(0.871)$ & -- & $0.723^{+0.027}_{-0.026}\,(0.724)$ & $2.967^{+0.518}_{-0.605}\,(3.041)$ & $1.497$ \\
$1$ & -- & $0.913^{+0.017}_{-0.018}\,(0.912)$ & -- & $0.626^{+0.027}_{-0.025}\,(0.630)$ & $2.310^{+0.621}_{-0.844}\,(2.473)$ & $1.378$ \\
$2$ & -- & $0.895^{+0.018}_{-0.019}\,(0.895)$ & -- & $0.702^{+0.029}_{-0.029}\,(0.705)$ & $2.434^{+0.650}_{-0.870}\,(2.590)$ & $1.001$ \\
\hline
\multicolumn{7}{c}{GG + SHAPE-derived IA fitting expanded by standard Legendre polynomials} \\
\hline
$0$ & 10.0   & $0.881^{+0.017}_{-0.017}\,(0.882)$ & $-0.200^{+0.006}_{-0.006}\,(-0.200)$ & $0.702^{+0.027}_{-0.026}\,(0.705)$ & $3.142^{+0.505}_{-0.600}\,(3.241)$ & $0.676$ \\
$0$ & 50.0   & $0.882^{+0.017}_{-0.017}\,(0.882)$ & $-0.173^{+0.005}_{-0.006}\,(-0.174)$ & $0.682^{+0.027}_{-0.026}\,(0.684)$ & $2.593^{+0.592}_{-0.739}\,(2.736)$ & $0.879$ \\
$0$ & 100.0 & $0.882^{+0.017}_{-0.018}\,(0.884)$ & $-0.157^{+0.006}_{-0.006}\,(-0.158)$ & $0.697^{+0.026}_{-0.026}\,(0.698)$ & $2.648^{+0.556}_{-0.676}\,(2.725)$ & $0.832$ \\
$1$ & 10.0   & $0.912^{+0.016}_{-0.017}\,(0.913)$ & $-0.187^{+0.005}_{-0.006}\,(-0.187)$ & $0.622^{+0.026}_{-0.025}\,(0.623)$ & $2.198^{+0.667}_{-0.876}\,(2.233)$ & $0.766$ \\
$1$ & 50.0   & $0.914^{+0.016}_{-0.016}\,(0.914)$ & $-0.163^{+0.005}_{-0.005}\,(-0.163)$ & $0.613^{+0.024}_{-0.023}\,(0.615)$ & $1.910^{+0.695}_{-0.958}\,(2.071)$ & $0.833$ \\
$1$ & 100.0 & $0.934^{+0.016}_{-0.016}\,(0.934)$ & $-0.142^{+0.005}_{-0.005}\,(-0.142)$ & $0.656^{+0.024}_{-0.024}\,(0.660)$ & $2.275^{+0.574}_{-0.779}\,(2.351)$ & $0.737$ \\
$2$ & 10.0   & $0.879^{+0.017}_{-0.017}\,(0.879)$ & $-0.197^{+0.006}_{-0.006}\,(-0.197)$ & $0.672^{+0.027}_{-0.026}\,(0.674)$ & $2.290^{+0.659}_{-0.889}\,(2.381)$ & $0.732$ \\
$2$ & 50.0   & $0.868^{+0.017}_{-0.018}\,(0.867)$ & $-0.169^{+0.006}_{-0.006}\,(-0.170)$ & $0.674^{+0.026}_{-0.025}\,(0.677)$ & $2.230^{+0.609}_{-0.802}\,(2.384)$ & $0.650$ \\
$2$ & 100.0 & $0.884^{+0.018}_{-0.018}\,(0.884)$ & $-0.145^{+0.005}_{-0.005}\,(-0.145)$ & $0.659^{+0.026}_{-0.024}\,(0.662)$ & $2.024^{+0.695}_{-0.957}\,(2.260)$ & $0.768$ \\
\hline
\multicolumn{7}{l}{\footnotesize$^a$ Projection depth is in unit of $h^{-1}$Mpc. }\\
\multicolumn{7}{l}{\footnotesize$^b$ Velocity dispersion is in unit of $h^{-1}$Mpc. }\\
\multicolumn{7}{l}{\footnotesize$^c$ Reduced $\chi^{2}$ values for the posterior mean and best-fitting results. The dof for the GG-only and the GG + SHAPE-derived}\\
\multicolumn{7}{l}{\footnotesize IA fitting expanded using a standard Legendre basis are $51$ and $212$, respectively. } \\
\end{tabular}
}
\end{table*}

\begin{table*}
\caption{Same as Table~\ref{tab:ia_fitting}, but the IA correlations are expanded using an associated Legendre polynomials. }
\label{tab:ia_fitting_assoc}
\resizebox{\textwidth}{!}{%
\begin{tabular}{clccccc}
\hline
\multicolumn{7}{c}{GG + SHAPE-derived IA fitting expanded by associated Legendre polynomials} \\
\hline
Realization & Projection$^a$ & $\bh$ & $\bK$ & $f$ & $\sigma_{v}$$^b$ & $\chi^{2}_{\rm total}$ / dof$^{c}$ \\
\hline
$0$ & 10.0  & $0.883^{+0.017}_{-0.017}\,(0.884)$ & $-0.200^{+0.006}_{-0.006}\,(-0.200)$ & $0.700^{+0.026}_{-0.026}\,(0.700)$ & $2.835^{+0.539}_{-0.642}\,(2.847)$ & $0.887$ \\
$0$ & 50.0  & $0.881^{+0.017}_{-0.017}\,(0.880)$ & $-0.174^{+0.006}_{-0.006}\,(-0.174)$ & $0.701^{+0.027}_{-0.027}\,(0.703)$ & $3.009^{+0.531}_{-0.653}\,(3.109)$ & $0.952$ \\
$0$ & 100.0 & $0.879^{+0.018}_{-0.018}\,(0.879)$ & $-0.160^{+0.006}_{-0.006}\,(-0.160)$ & $0.706^{+0.026}_{-0.026}\,(0.708)$ & $2.989^{+0.521}_{-0.612}\,(3.071)$ & $0.950$ \\
$1$ & 10.0  & $0.908^{+0.016}_{-0.017}\,(0.907)$ & $-0.189^{+0.005}_{-0.006}\,(-0.189)$ & $0.622^{+0.026}_{-0.024}\,(0.625)$ & $2.085^{+0.658}_{-0.915}\,(2.251)$ & $0.900$ \\
$1$ & 50.0  & $0.902^{+0.017}_{-0.017}\,(0.903)$ & $-0.160^{+0.005}_{-0.005}\,(-0.160)$ & $0.630^{+0.026}_{-0.024}\,(0.633)$ & $2.086^{+0.662}_{-0.919}\,(2.283)$ & $0.989$ \\
$1$ & 100.0 & $0.904^{+0.016}_{-0.017}\,(0.903)$ & $-0.146^{+0.005}_{-0.005}\,(-0.146)$ & $0.636^{+0.026}_{-0.026}\,(0.640)$ & $2.441^{+0.609}_{-0.774}\,(2.612)$ & $0.836$ \\
$2$ & 10.0  & $0.891^{+0.017}_{-0.017}\,(0.889)$ & $-0.195^{+0.006}_{-0.006}\,(-0.196)$ & $0.662^{+0.029}_{-0.028}\,(0.670)$ & $2.410^{+0.680}_{-0.856}\,(2.691)$ & $0.794$ \\
$2$ & 50.0  & $0.882^{+0.018}_{-0.018}\,(0.882)$ & $-0.169^{+0.006}_{-0.006}\,(-0.169)$ & $0.662^{+0.026}_{-0.025}\,(0.667)$ & $2.173^{+0.669}_{-0.931}\,(2.364)$ & $0.831$ \\
$2$ & 100.0 & $0.908^{+0.017}_{-0.018}\,(0.907)$ & $-0.150^{+0.006}_{-0.005}\,(-0.151)$ & $0.690^{+0.027}_{-0.026}\,(0.694)$ & $2.213^{+0.642}_{-0.900}\,(2.412)$ & $0.889$ \\
\hline
\multicolumn{7}{l}{\footnotesize$^a$ Projection depth is in unit of $h^{-1}$Mpc. }\\
\multicolumn{7}{l}{\footnotesize$^b$ Velocity dispersion is in unit of $h^{-1}$Mpc. }\\
\multicolumn{7}{l}{\footnotesize$^c$ Reduced $\chi^{2}$ values for the posterior mean and best-fitting results. The dof for the GG + SHAPE-derived IA fitting using}\\
\multicolumn{7}{l}{\footnotesize an associated Legendre basis is $140$. } \\
\end{tabular}
}
\end{table*}

\begin{figure}
\includegraphics[width=\columnwidth]{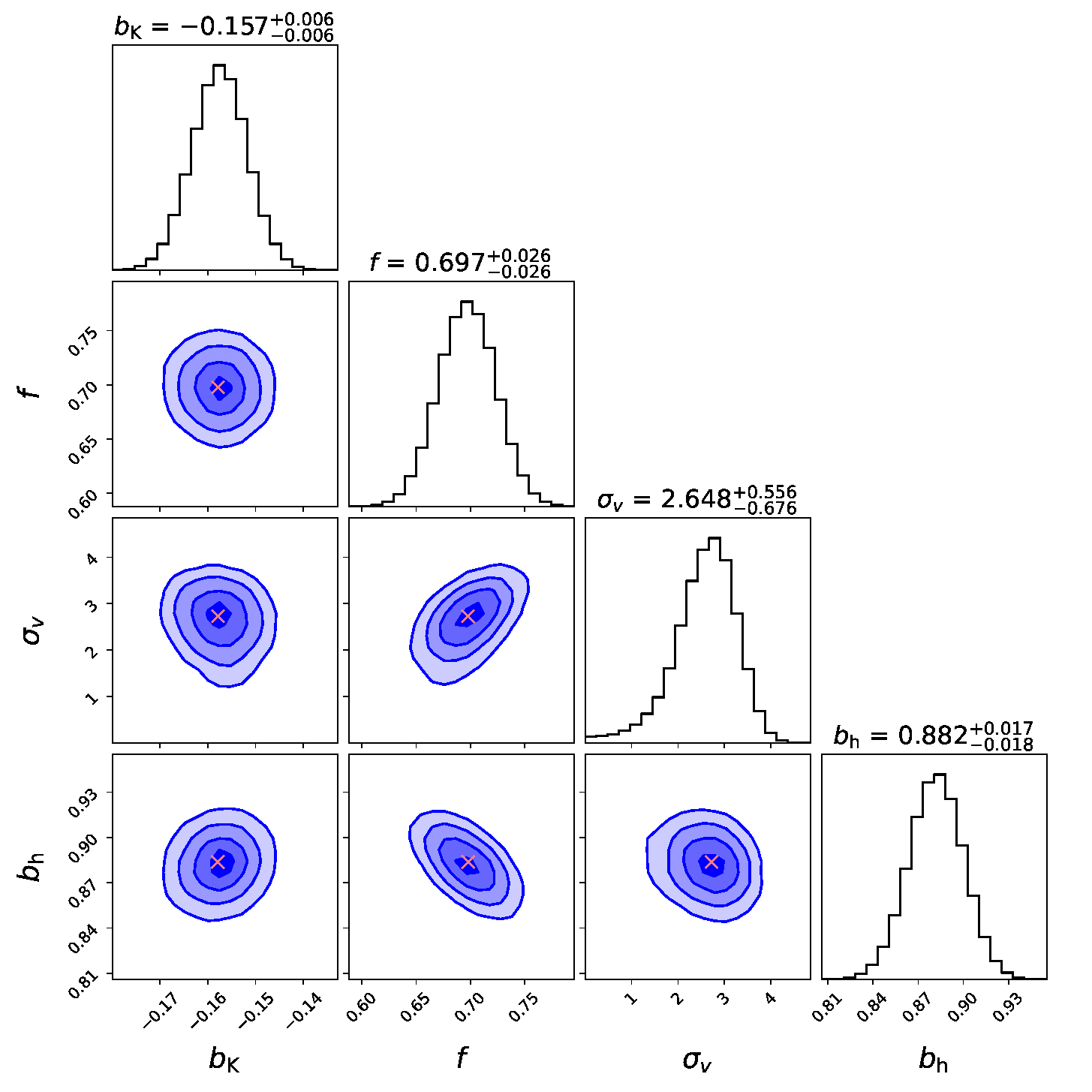}
\caption{Marginalized 1D posterior distributions and 2D probability contours obtained from the fitting procedure of the GG and SHAPE-derived IA correlations with $100.0$~$h^{-1}$Mpc projection in redshift space. The contours represent $68\%$, $95\%$, and $99.7\%$ confidence levels and orange crosses on the contours show the best-fitting values during the MCMC procedure. At $z=0.319508$, the fiducial value of the linear growth rate is $f=0.6936$ based on the Planck~2015 cosmological parameters. }
\label{fig:100mpc_contour}
\end{figure}

The error contours obtained from the parameter inference using the GG and SHAPE-derived IA correlations with a $100\,h^{-1}\mathrm{Mpc}$ projection depth for Realization~0 are presented in Figure~\ref{fig:100mpc_contour}. 
The linear growth rate is effectively constrained to $f = 0.697^{+0.026}_{-0.026}$ through the joint fitting of GG and IA correlations, along with simultaneous constraints on the halo bias, shape bias, and velocity dispersion parameters. 
This measured $f$ value closely aligns within the $1\sigma$ uncertainty range of the fiducial $\Lambda$CDM cosmological prediction ($f=0.6936$ at $z=0.319508$). 

Hence, these findings reinforce the SHAPE technique as a promising approach for accurate cosmological inference from IA signals. 
The consistency with the fiducial cosmology at larger projection depths observed here and in other realizations supports the robustness of our method across different observational scenarios. 

Given these promising results, future studies should extend the SHAPE technique to real observational datasets from large-scale galaxy surveys. 
Further exploration could also examine its capability to constrain additional cosmological parameters such as the Hubble parameter, geometric distances, and the dark energy equation of state \citep[e.g.,][]{johnston19,taruya20,okumura22,shim24a,shim24b}. 
Leveraging the dual sensitivity of IA signals to both density and tidal fields will enable more comprehensive cosmological analyses and tighter parameter constraints. 

\section{DISCUSSION} \label{sec:discussion}
\subsection{Expanding using Legendre basis versus associated Legendre basis} \label{subsec:basis_comparison}
Recent studies have shown that the GI and II~(-) correlations can also be expanded using the associated Legendre polynomials \citep{kurita22,kurita23,okumura24,inoue25}, which provide an alternative basis incorporating additional angular information compared to the standard Legendre expansion (equation~\ref{eq:legendre_estimator}). 
In this section, we investigate whether the associated Legendre polynomial basis offers improved accuracy or robustness in IA correlation modelling. 

The GI and II~(-) correlations computed from the $N$-body simulations can be expanded using the associated Legendre polynomials $L_{\ell}^{m}$ as: 
\begin{equation} 
\xi_{{\rm Y}, \ell m} (r) = \frac{(2\ell + 1)}{2} \frac{(\ell-m)!}{(\ell+m)!} \sum_{i, j} \xi_{\rm Y}(r_{ij}) L_{\ell}^{m} (\mu_{ij}),
\label{eq:assoc_legendre_estimator} 
\end{equation}
where ${\rm Y}$ denotes the GI or II~(-) correlations. 
We focus only on the dominant terms, $\ell = m = 2$ for the GI correlation and $\ell = m = 4$ for the II~(-) correlation, as higher-order contributions are negligible. 

The MCMC fitting follows the same procedure described in Section~\ref{subsec:model_fitting}, with the likelihood estimation (equation~\ref{eq:fitting_model}) adjusted to include the associated Legendre polynomial components: 
\begin{equation}
\begin{aligned}
\mathrm{X} &= \mathrm{X}' \\
            &= \{\,
               \mathrm{hh},\;
               \mathrm{h+}(\ell=m=2),\\
            &\quad
               \mathrm{+},\;
               \mathrm{-}(\ell=m=4)
             \}\,.
\end{aligned}
\end{equation}
All other settings, including jackknife resampling and MCMC sampling, remain identical. 

\begin{figure}
\includegraphics[width=\columnwidth]{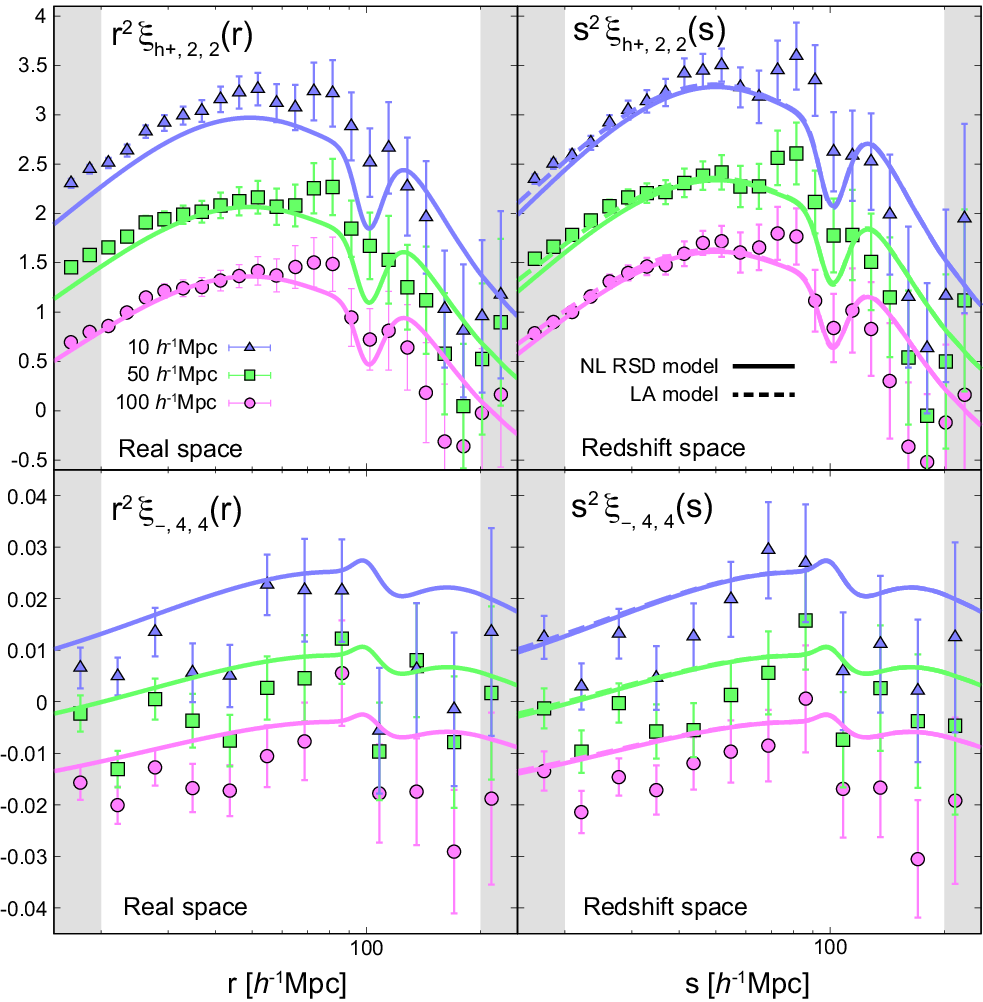}
\caption{The GI (top panels) and II~(-) (bottom panels) correlation functions expanded using the associated Legendre polynomial basis, measured in both real space (left) and redshift space (right). Different colors indicate different projection depths along the line-of-sight: $10.0\,h^{-1}{\rm Mpc}$ (upward triangles), $50.0\,h^{-1}{\rm Mpc}$ (squares), and $100.0\,h^{-1}{\rm Mpc}$ (circles). To facilitate visual comparison, constant vertical offsets are added to the correlation functions: for the GI correlations, offsets of $-0.5$ and $-1.0$ are applied to the $50.0$ and $100.0\,h^{-1}{\rm Mpc}$ projections, respectively; for the II~(-) correlations, offsets of $-0.01$ and $-0.02$ are similarly applied. Solid and dashed curves correspond to the posterior mean fits obtained from the NL RSD and the LA models, respectively. Shaded grey regions indicate scales excluded from the fitting analysis (${\rm s}<20\,h^{-1}{\rm Mpc}$ and ${\rm s}>200\,h^{-1}{\rm Mpc}$).
}
\label{fig:associated}
\end{figure}

Figure~\ref{fig:associated} presents the fitting results of the GI and II~(-) correlations expanded by the associated Legendre polynomial basis for each projection depth, with fitting results summarized in Table~\ref{tab:ia_fitting_assoc}. 
Similar to Figure~\ref{fig:gi-ii}, the IA correlations from the simulations are accurately represented by the NL RSD model in redshift space. 
The GI correlations obtained using the associated Legendre basis clearly reveal the BAO-scale feature around $100\,h^{-1}{\rm Mpc}$, similar to those obtained using the standard Legendre polynomial approach. 
Moreover, the BAO trough becomes more pronounced when using larger projection depths (e.g., $100\,h^{-1}{\rm Mpc}$) compared to shallower projection depths. 
While these observed features in the GI correlation are suggestive of BAO signals, precise confirmation would benefit from further analysis involving detailed model comparisons. 
On the other hand, the II~(-) correlations expanded by the associated Legendre basis show consistency with theoretical predictions, albeit with lower measured amplitudes. 
Although the theoretical model predictions exhibit clear BAO signatures, the measured II~(-) correlations have larger uncertainties, making it challenging to conclusively confirm BAO features from the II~(-) correlations alone. 

Comparing the results obtained from the two polynomial bases, we find that cosmological parameter constraints, particularly the linear growth rate parameter $f$, are modestly improved when using the associated Legendre polynomial expansion. 
For instance, in Realization~0 with a projection depth of $10\,h^{-1}{\rm Mpc}$, the associated Legendre polynomial basis yields $f = 0.700^{+0.026}_{-0.026}$, which is slightly closer to the fiducial value $f = 0.6936$ compared to the result from the standard Legendre basis ($f = 0.702^{+0.027}_{-0.026}$). 
Similar modest improvements in parameter accuracy are consistently observed for larger projection depths and across other realizations. 

More significantly, the associated Legendre polynomial basis consistently produces reduced $\chi^2$ values closer to unity across various projection depths and realizations (ranging approximately from $0.794$ to $0.989$). 
In contrast, the standard Legendre basis occasionally yields slightly lower $\chi^2$ values, potentially indicating mild over-fitting due to the capture of statistical noise. 

Taken together, these findings demonstrate that the associated Legendre polynomial basis offers a more imporoved statistically significance representation of IA correlations. 
While improvements in the accuracy of the growth rate parameter are modest compared to the standard Legendre polynomial basis, the associated Legendre polynomial basis substantially reduces potential over-fitting, enhancing the reliability and robustness of cosmological parameter constraints. 
Therefore, despite the modest gain in parameter precision, adopting the associated Legendre polynomial basis remains clearly advantageous, especially for future IA analyses focused on accurate and reliable cosmological inference. 

\subsection{BAO trough measured by the SHAPE technique} \label{subsec:bao_scale}
The detection of the BAO signal in large-scale structure surveys serves as a critical standard ruler for cosmological distance measurements. 
While BAO features have been extensively studied in galaxy clustering analyses, their imprint on IA correlations, particularly the GI correlation, offers an additional avenue for cosmological investigation. 
Recent work by \citet{xu23} demonstrated a $2-3\sigma$ detection of the BAO trough in the observed GI correlation function, combining the CMASS galaxy sample from the SDSS Baryon Oscillation Spectroscopic Survey (BOSS) and shape measurements from the DESI Legacy Imaging Surveys. 
Their analysis constrained the distance parameter $D_V/r_{\mathrm{d}}$ at redshift $0.57$ to a precision of $3-5\%$, highlighting the complementary role of GI correlations, which reduced uncertainty by approximately $10\%$ compared to GG correlations alone. 

\begin{figure}  
\includegraphics[width=\columnwidth]{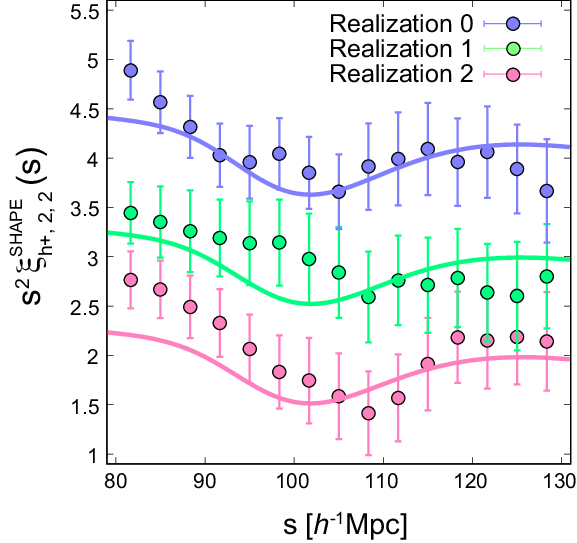}  
\caption{The GI correlation function in redshift space expanded using the associated Legendre polynomial basis, $\xi^{\rm SHAPE}_{\rm h+, 2, 2}(s)$, in linear scale measured from the shape catalogues with a $100\,h^{-1}$Mpc projection for each realization. To facilitate visual comparison, constant offsets are added to the correlation functions: $+2$ for Realization~0 and $+1$ for Realization~1. Solid curves represent the posterior mean results from the model fits using the NL RSD model described in Section~\ref{sec:results}. The BAO trough is successfully captured in all SHAPE-based measurements, although slight shifts toward larger scales are observed compared to the corresponding model predictions.}
\label{fig:bao_sia}  
\end{figure}

In our study, we employ the SHAPE technique to measure the GI correlation function and assess the consistency of the observed BAO scale with theoretical predictions. 
Figure~\ref{fig:bao_sia} presents the GI correlation function in redshift space, expanded using the associated Legendre polynomial basis, $\xi^{\rm SHAPE}_{\rm h+, 2, 2}(s)$, in linear scale derived from shape catalogues with a $100\,h^{-1}$Mpc projection in each of our three simulation realizations, and compares them with the posterior mean results from the model fitting using the NL RSD model given in Section~\ref{sec:results}. 
Our results indicate that the SHAPE-based GI correlations consistently capture the BAO trough at scales broadly consistent with theoretical predictions across all three realizations, reinforcing the robustness of the method. 
However, slight shifts of the trough towards larger scales are observed relative to the model predictions. 
Such shifts may arise from several factors. 
Firstly, the SHAPE method estimates halo shapes from projected subhalo distributions, and associated projection effects can introduce shape noise and subtle phase shifts in the measured BAO feature. 
Secondly, the finite size ($1\,h^{-1}$Gpc) of our simulation boxes introduces sample variance, potentially affecting the precise determination of the BAO trough position. 
Moreover, periodic boundary conditions could induce subtle large-scale clustering artifacts, marginally influencing the BAO scale measured by the SHAPE method. 
A more comprehensive analysis incorporating multiple realizations with larger simulation boxes would be necessary to robustly quantify these effects. 

Despite these minor deviations, the consistent recovery of the BAO-scale feature across multiple realizations supports the effectiveness of the SHAPE approach in capturing large-scale tidal alignment signals, although a definitive confirmation of the BAO trough would benefit from more detailed quantitative analyses. 
Our findings, coupled with the observational results of \citet{xu23}, underscore the promising potential of IA measurements, particularly the GI correlation, as complementary cosmological probes. 
Applying the SHAPE approach to forthcoming large-scale observational surveys can offer valuable constraints on fundamental cosmological parameters, enhancing the precision of future cosmological analyses. 

\section{Conclusions, summary, and future prospects} \label{sec:conclusions}
In this paper, we proposed a novel method termed the Subhalo-based Halo Alignment and Projected Ellipticity (SHAPE) approach, which measures IA correlations of cluster-scale haloes using the spatial distributions of their member subhaloes. 
Unlike conventional IA analyses that rely upon precise measurements of individual galaxy shapes, the SHAPE technique statistically infers the overall shapes and orientations of cluster haloes from their subhalo populations. 
This significantly mitigates observational uncertainties associated with galaxy shape measurements, thereby broadening the applicability of IA analyses across diverse observational datasets. 
Validated using high-resolution $N$-body simulations, the SHAPE approach accurately recovered halo shapes and orientations, providing IA correlations consistent with direct particle-based measurements. 

The major findings of this work are summarized as follows:

\begin{enumerate}

\item The SHAPE method robustly measures halo shapes from subhalo distributions. 
While large projection depths ($\gtrsim100\,h^{-1}{\rm Mpc}$) introduce contamination in shape measurements, moderate projection depths ($10-100$ $h^{-1}{\rm Mpc}$) effectively recover intrinsic shapes and orientations even in the presence of the Fingers-of-God effect (Figure~\ref{fig:comp_shape}). 

\item Using simulated cluster haloes, we calculated the GG, GI, II~(+), and II~(-) correlation functions based on SHAPE-derived halo shapes (Figure~\ref{fig:gi-ii}). 
These anisotropic correlations were expanded into multipoles using Legendre and associated Legendre polynomial bases. 
The analytical model of the non-linear correlations with the redshift-space distortion effect (NL RSD model) successfully reproduced both the measured GG and IA correlations, confirming the validity of the SHAPE technique. 

\item The GI correlation measured with the SHAPE method effectively captures the BAO-scale feature at $\sim100\,h^{-1}{\rm Mpc}$, particularly at larger projection depths ($100\,h^{-1}{\rm Mpc}$). 
In contrast, shallow projection depths ($\sim 10\,h^{-1}{\rm Mpc}$) yield a weaker detection of this feature, as they primarily capture subhaloes near cluster centers and underestimate contributions from the cluster outskirts, which play a crucial role in aligning haloes with the large-scale tidal field. 
Hence, deeper projection depths enhance the fidelity of the measured large-scale tidal alignment, improving the detection of the BAO trough despite increased foreground/background contamination. 

\item Compared to the GG-only analysis, jointly fitting GG and SHAPE-derived IA correlations substantially improved both the accuracy and precision of the linear growth rate measurement. 
Specifically, deviations from the Planck~2015 fiducial value ($f=0.6936$) were notably reduced (e.g., from approximately $4.2\%$ to $0.5\%$ in Realization0), accompanied by modest reductions in statistical uncertainties across all realizations. 
This highlights the benefit of incorporating SHAPE-derived IA correlations into cosmological analyses. 

\item GI and II~(-) correlations expanded using the associated Legendre polynomial basis provided modest yet consistent improvements in statistical significance compared to the standard Legendre basis (Figure~\ref{fig:associated} and Table~\ref{tab:ia_fitting} and \ref{tab:ia_fitting_assoc}). 
Although the inferred cosmic linear growth rate parameter, $f$, showed only slight improvements towards the Planck~2015 fiducial value ($f = 0.6936$), the reduced $\chi^2$ values consistently approached unity across all realizations, indicating a more statistically balanced fit. 
These subtle improvements highlight the reliability and effectiveness of the associated Legendre polynomial basis for IA-based cosmological analyses. 

\end{enumerate}

Although the measured II correlations exhibit relatively large uncertainties due to the limited number of cluster haloes, the overall amplitudes of the monopole component align closely with theoretical predictions from the NL RSD model. 
While the data suggest hints of BAO-scale features in the II correlations, definitive confirmation would require improved statistical precision from larger simulations. 
Notably, including the II correlations in joint fitting modestly improves constraints on the linear growth rate parameter, underscoring their value in cosmological analyses. 
Future analyses using larger simulation volumes or large observational datasets will provide more precise validations of these correlations and further clarify their cosmological implications. 

Our indication of BAO signatures obtained using the SHAPE technique, combined with recent observational results, demonstrates its potential for precise cosmological inference from extensive galaxy surveys. 
The pronounced IA signals from cluster-scale haloes, when combined with galaxy clustering measurements, can provide stringent constraints on fundamental cosmological parameters such as the Hubble parameter and the linear growth rate \citep{taruya20,okumura22,okumura23}. 
Applying the SHAPE approach to existing and upcoming large-scale observational surveys, including SDSS BOSS, HSC SSP \citep{aihara18}, Rubin LSST \citep{ivezic19}, Euclid \citep{euclid22}, Subaru PFS \citep{takada14}, and DESI \citep{desi16}, represents a crucial next step toward achieving more robust cosmological constraints. 
In preparation for these observational applications, analyses based on realistic mock catalogues constructed from halo occupation distributions calibrated to high-redshift data \citep[e.g.,][]{ishikawa24,ishikawa25} will be essential to refine the SHAPE method and validate its performance under realistic observational conditions. 

\begin{acknowledgments}
We thank Masahiro Takada, Jingjing Shi, Jounghun Lee, Tomoaki Ishiyama, Toshiki Kurita, Hironao Miyatake, and Ken Osato for giving us useful comments and discussion for this study. 

SI acknowledges support from ISHIZUE 2022 of Kyoto University. 
This work was supported in part by MEXT/JSPS KAKENHI Grant Numbers JP23K13145 (SI), JP20H05861 and JP23K20844 (AT and TN), JP23K25868 (AT), JP22K03634, JP24H00215 (TN), and JP24H00221 (TN and ST). 
TO acknowledges the support of the Taiwan National Science and Technology Council under Grants No. NSTC 112-2112-M-001034- and No. NSTC 113-2112-M-001-011- and the Investigator Project Grant, Academia Sinica (AS-IV-114-M03) for the period of 2025-2029. 

The authors thank the Yukawa Institute for Theoretical Physics at Kyoto University. 
Discussions during the YITP workshop YITP-W-22-16 on ``New Frontiers in Cosmology with the Intrinsic Alignments of Galaxies'' were useful to complete this work. 

Numerical computations were carried out on Cray XC50 and analysis servers at Center for Computational Astrophysics, National Astronomical Observatory of Japan. 
Data analysis was in part carried out on the Multi-wavelength and the Large-scale data analysis system co-operated by the Astronomy Data Center (ADC) and Subaru Telescope, National Astronomical Observatory of Japan. 
\end{acknowledgments}

\bibliography{ref}{}
\end{document}